\documentclass[usenatbib,usegraphicx]{mn2e}

\usepackage{longtable}

\def\CIVdbl{{\rm C~}\kern 0.1em{\sc iv}~$\lambda\lambda 1548, 1550$}
\def\MgIIdbl{{\rm Mg~}\kern 0.1em{\sc ii}~$\lambda\lambda 2796, 2803$}
\def\NVdbl{{\rm N}\kern 0.1em{\sc v}~$\lambda\lambda 1238, 1242$}  
\def\OVIdbl{{\rm O}\kern 0.1em{\sc vi}~$\lambda\lambda 1031, 1037$}
\def\SiIVdbl{{\rm Si~}\kern 0.1em{\sc iv}~$\lambda\lambda1394, 1403$}
\def\AlIIIdbl{{\rm Al~}\kern 0.1em{\sc iii}~$\lambda\lambda1855,1863$}
\def\FeIIdbl{{\rm Fe~}\kern 0.1em{\sc ii}~$\lambda\lambda 2383, 2600$}

\def\AlII{\hbox{{\rm Al~}\kern 0.1em{\sc ii}}}
\def\AlIII{\hbox{{\rm Al~}\kern 0.1em{\sc iii}}}
\def\CaI{\hbox{{\rm Ca}\kern 0.1em{\sc i}}}
\def\CaII{\hbox{{\rm Ca}\kern 0.1em{\sc ii}}}
\def\CrII{\hbox{{\rm Cr}\kern 0.1em{\sc ii}}}
\def\CII{\hbox{{\rm C~}\kern 0.1em{\sc ii}}}
\def\CIII{\hbox{{\rm C~}\kern 0.1em{\sc iii}}}
\def\CIV{\hbox{{\rm C~}\kern 0.1em{\sc iv}}}
\def\CV{\hbox{{\rm C}\kern 0.1em{\sc v}}}

\def\HI{\hbox{{\rm H~}\kern 0.1em{\sc i}}}
\def\HII{\hbox{{\rm H~}\kern 0.1em{\sc ii}}}
\def\Lya{\hbox{{\rm Ly}\kern 0.1em$\alpha$}}
\def\Lyb{\hbox{{\rm Ly}\kern 0.1em$\beta$}}
\def\Lyg{\hbox{{\rm Ly}\kern 0.1em$\gamma$}}
\def\Lyfive{\hbox{{\rm Ly}\kern 0.1em$5$}}
\def\Lysix{\hbox{{\rm Ly}\kern 0.1em$6$}}
\def\Lyseven{\hbox{{\rm Ly}\kern 0.1em$7$}}
\def\Lyeight{\hbox{{\rm Ly}\kern 0.1em$8$}}
\def\Lynine{\hbox{{\rm Ly}\kern 0.1em$9$}}
\def\Lyten{\hbox{{\rm Ly}\kern 0.1em$10$}}
\def\HeI{\hbox{{\rm He}\kern 0.1em{\sc i}}}
\def\HeII{\hbox{{\rm He}\kern 0.1em{\sc ii}}}
\def\FeI{\hbox{{\rm Fe~}\kern 0.1em{\sc i}}}
\def\FeII{\hbox{{\rm Fe~}\kern 0.1em{\sc ii}}}
\def\FeIII{\hbox{{\rm Fe~}\kern 0.1em{\sc iii}}}
\def\MnII{\hbox{{\rm Mn}\kern 0.1em{\sc ii}}}
\def\MgI{\hbox{{\rm Mg~}\kern 0.1em{\sc i}}}
\def\MgII{\hbox{{\rm Mg~}\kern 0.1em{\sc ii}}}
\def\MgIII{\hbox{{\rm Mg~}\kern 0.1em{\sc iii}}}
\def\MgIV{\hbox{{\rm Mg~}\kern 0.1em{\sc iv}}}
\def\NaI{\hbox{{\rm Na}\kern 0.1em{\sc i}}}
\def\NV{\hbox{{\rm N}\kern 0.1em{\sc v}}}
\def\NII{\hbox{{\rm N}\kern 0.1em{\sc ii}}}
\def\NIII{\hbox{{\rm N}\kern 0.1em{\sc iii}}}

\def\Mg{\hbox{{\rm Mg}}}
\def\Fe{\hbox{{\rm Fe}}}
\def\Zn{\hbox{{\rm Zn}}}
\def\OVI{\hbox{{\rm O}\kern 0.1em{\sc vi}}}
\def\OII{\hbox{[{\rm O}\kern 0.1em{\sc ii}]}}
\def\OIII{\hbox{[{\rm O}\kern 0.1em{\sc iii}]}}
\def\SiII{\hbox{{\rm Si~}\kern 0.1em{\sc ii}}}
\def\SiIII{\hbox{{\rm Si~}\kern 0.1em{\sc iii}}}
\def\SiIV{\hbox{{\rm Si~}\kern 0.1em{\sc iv}}}
\def\SII{\hbox{{\rm S}\kern 0.1em{\sc ii}}}
\def\SIII{\hbox{{\rm S}\kern 0.1em{\sc iii}}}
\def\SIV{\hbox{{\rm S}\kern 0.1em{\sc iv}}}
\def\TiII{\hbox{{\rm Ti}\kern 0.1em{\sc ii}}}
\def\ZnII{\hbox{{\rm Zn}\kern 0.1em{\sc ii}}}
\def\kms{\hbox{km~s$^{-1}$}}      

\def\cc{\hbox{cm$^{-3}$}}

\def\a{$\alpha$ }
\def\s{$\sigma$ }
\def\lb{$\lambda$}
\def\lsim{\mathrel{\rlap{\lower4pt\hbox{\hskip1pt$\sim$}}
    \raise1pt\hbox{$<$}}}                
\def\gsim{\mathrel{\rlap{\lower4pt\hbox{\hskip1pt$\sim$}}
    \raise1pt\hbox{$>$}}}                

\title[]{Evolution of the Population of Very Strong {\MgII} Absorbers}

\author[P.~Rodr\'iguez Hidalgo et al.]
{\parbox{\textwidth}{Paola~Rodr\'iguez~Hidalgo$^{1,2}$\thanks{E-mail: \texttt{prh@yorku.ca}}
Kaylan~Wessels$^{1,3}$,
Jane~C.~Charlton$^{1}$,
Anand~Narayanan$^{1,4}$,
Andrew~Mshar$^{1}$,
Antonino Cucchiara$^{1,5}$, and
Therese~Jones$^{1,6}$}\vspace{0.4cm} \\
\parbox{\textwidth}{$^{1}$Department of Astronomy \& Astrophysics, Pennsylvania State University, University Park, PA 16802 \\
$^{2}$Department of Physics \& Astronomy, York University, Toronto, ON, Canada M3J 1P3 \\
$^{3}$Materials Department, University of California, Santa Barbara, CA 93106 \\
$^{4}$Department of Earth \& Space Sciences, Indian Institute of Space Science \& Technology, Thiruvananthapuram, India 695547\\
$^{5}$Department of Astronomy and Astrophysics \& UCO/Lick Observatory, University of California, 1156 High Street, Santa Cruz, CA 95064\\
$^{6}$Department of Astronomy, University of California, B-20 Hearst Field Annex, Berkeley, CA 94720
}}

\begin{document}

\pagerange{\pageref{firstpage}--\pageref{lastpage}} \pubyear{2011}
\maketitle
\label{firstpage}

\begin{abstract} 

We present a study of the evolution of several classes of {\MgII}
absorbers, and their corresponding {\FeII} absorption, over a large
fraction of cosmic history: 2.3 to 8.7 Gyrs from the Big Bang. Our
sample consists of 87 strong ($W_r(\MgII)>$~0.3 \AA) {\MgII}
absorbers, with redshifts 0.2~$ < z <$~2.5, measured in 81 quasar
spectra obtained from the Very Large Telescope (VLT) / Ultraviolet and
Visual Echelle Spectrograph (UVES) archives of high-resolution spectra
(R $\sim$ 45,000). No evolutionary trend in $W_r(\FeII)/W_r(\MgII)$ is found for
moderately strong {\MgII} absorbers (0.3~$<W_r(\MgII)<$~1.0
{\AA}). However, at lower redshifts we find an absence of very strong
{\MgII} absorbers (those with $W_r(\MgII)>$~1 {\AA}) with small ratios
of equivalent widths of {\FeII} to {\MgII}.  At high redshifts,
very strong {\MgII} absorbers with both small and large $W_r(\FeII)/W_r(\MgII)$ values
are present. We compare our findings to
a sample of 100 weak {\MgII} absorbers
($W_r(\MgII) <$~0.3 \AA) found in the same quasar spectra
by Narayanan et~al. (2007).

The main effect driving the evolution of very strong {\MgII} systems
is the difference between the kinematic profiles at low and high
redshift. At high redshift, we observe that, among the very strong
{\MgII} absorbers, all of the systems with small ratios of
$W_r(\FeII)/W_r(\MgII)$ have relatively large velocity spreads, resulting
in less saturated profiles.
At low redshift, such  kinematically spread systems are absent,
and both {\FeII} and {\MgII} are saturated, leading to
$W_r(\FeII)/W_r(\MgII)$ values that are all close to 1.
The high redshift, small $W_r(\FeII)/W_r(\MgII)$ systems
could correspond to sub-DLA systems, many of which have large velocity
spreads and are possibly linked to superwinds in star forming
galaxies.  In addition to the change in saturation due to kinematic
evolution, the smaller $W_r(\FeII)/W_r(\MgII)$ values could be
due to a lower abundance of {\Fe} at high redshifts, which would
indicate relatively early stages of star formation in those
environments.

\end{abstract}

\section{Introduction}
\label{sec:1}

Quasar absorption lines allow the study of galaxies and their halos
with no bias toward specific environments because of the random
distribution of lines of sight in the sky. In particular, high
resolution studies of quasar absorption lines help us better
characterize the kinematic properties and chemical content of the
absorbing gas. The kinematic composition of individual galaxies can be
studied statistically with quasar absorption lines
(\citealt{Charlton98}). Profile shapes of the absorption lines show
clouds of gas at different velocities relative to galaxy centroids,
establishing, for example, whether the absorption is consistent with
absorption in the disk and/or the extended halo (\citealt{Steidel02b};
\citealt{Kacprzak11}). Also, understanding the nature and evolution of
galaxies requires assessing their star formation history, which can be
traced by their metal content (e.g., \citealt{Matteucci08} and
references therein). Although the number of massive stars is smaller
than the number of less massive stars, the former have a larger impact
on the galaxy's chemical enrichment through supernovae
explosions. Different types of supernovae (SNe) contribute differently
to this enrichment: core-collapse SNe (Type II) produce a larger
contribution of $\alpha$--elements (such as \Mg) relative to \Fe\,
within the first billion years of the formation of a stellar
population. \Fe\, is mostly generated in a later phase by Type Ia
SNe. Thus, the ratio of \Fe\, relative to \Mg\, may be used as a
measure of the stage in the star formation history of a galaxy. Quasar
absorption line studies are a probe of this kind of information.

Among the absorption features that can be used in extragalactic
studies, the resonance doublet transitions of one $\alpha$--element,
\MgII\,\lb\lb2796,2803, are well suited due to their strength and
their accessibility at
optical wavelengths from redshifts of 0.2 up to 2.7, covering a large
range in cosmic ages (from approx. 2.5 to 11.3 Gyrs). These
intervening {\MgII} absorption lines have been historically (and
originally arbitrarily) classified based on the strength of their
rest-frame equivalent width in the 2796~{\AA} transition
($W_r({\MgII})$\footnote{Throughout the paper we use $W_r(\MgII)$ to describe
the rest-frame equivalent width of the {\MgII} \lb2796 transition}).
Posteriorly, physical properties and origins for these systems have defined
statistically distinct populations, although the equivalent width
boundaries between these different populations are not sharp.  Strong
{\MgII} absorption systems are those with $W_r({\MgII}) \geq$~0.3 \AA, while
weak \MgII\, absorption is defined to have $W_r({\MgII}) <$~0.3 \AA. Several
sub-classifications have been proposed among the strong absorption
systems: very strong absorbers are those with $W_r({\MgII}) \geq$~1
\AA, and the term ``ultrastrong'' is reserved for those \MgII\,
absorption systems with $W_r({\MgII}) \gsim$~3 \AA\, (\citealt{Nestor05};
\citealt{Nestor07}).  Strong \MgII\, absorption has generally been
found to be connected to galaxies. Luminous galaxies
($\sim0.1-5{L_\star}$ galaxy) near quasar lines of sight show strong
\MgII\, absorption within 60~kpc in $\approx$~75\% of all cases
(\citealt{Bergeron92}; \citealt{Steidel95}; \citealt{Steidel97};
\citealt{Churchill05}; \citealt{Zibetti07}; \citealt{Chen10}; \citealt{Nestor11}; \citealt{Rao11}).  
Recent work has suggested
that the {\MgII} halos are patchy, with $\sim$50\% covering fraction
for {\MgII} (\citealt{Tripp05}; \citealt{Churchill07};
\citealt{Kacprzak08}), and it seems this covering fraction might be
even smaller at smaller redshifts (less than 40\% for strong {\MgII}
absorbers at $z \sim$~0.1; \citealt{Barton09}).

Strong {\MgII} absorbers have been found to be good tracers of
cold/warm and low-ionization gas.  \citet{Rao06} suggests that
$\approx$~80\% of very strong \MgII\, absorption traces damped
Lyman-$\alpha$ (DLA) galaxies ($N(\HI)>10^{20.3}$), but the relation
between $N(\HI)$ and $N(\MgII)$ does not prevail for the largest
$W_r(\MgII)$. \citet{Nestor07} indicates that ``while it is likely that
a large fraction of ultrastrong MgII absorption (perhaps~$>$~60\%)
have $N(\HI)>$ 10$^{20.3}$ atoms cm$^{-2}$, most (perhaps~$\sim$~90\%)
DLAs have $W_r(\MgII) \lsim$~3 {\AA}''. Indeed, $W_r(\MgII)$ and
$N(\HI)$ do not correlate as much in this regime and ultrastrong
{\MgII} systems can be found in sub--DLA galaxies (10$^{19}<N(\HI)<$
10$^{20.3}$ atoms cm$^{-2}$) as well (\citealt{Rao11}). Moreover,
\citet{Kulkarni10} has found that the {\MgII} associated with sub-DLAs
tends to show larger velocity spreads on average than that associated
with DLAs. Together with the finding that sub-DLAs are more metal-rich
than DLAs, \citet{Kulkarni10} suggests that the difference is due to
their star formation histories, where galaxies associated with
sub-DLAs would tend to be more massive, and suffer faster star
formation and gas consumption, which would result in less $N(\HI)$ and
more metal production. If the star formation has occurred recently, we
would expect to see kinematic signatures of it in the {\MgII}
absorption profiles, due to outflows, also called superbubbles or
superwinds.  Moreover, we would expect evolution of very strong
{\MgII} absorption profiles as the star formation rate in galaxies
shows a decline at redshifts lower than $z \sim 1$.

Indeed, {\MgII} absorbers have been observed to show evolution with
redshift. In a survey of $\sim$~3700 quasars from an early data release
of Sloan Digital Sky Survey (SDSS), \citet{Nestor05} found that the
number of very strong {\MgII} absorbers ($W_r > 1$ \AA) was not
consistent with the expectations for cosmological non--evolution,
showing larger numbers at high redshift ($z \gsim$~1.2). This trend is
most significant as $W_r(\MgII)$ increases. \citet{Prochter06}
confirmed these results for the incidence of {\MgII} in a study of
45,023 quasars from the Data Release 3 (DR3) of SDSS. The decline in
number of very strong {\MgII} absorbers with time is consistent with
the decline of cosmic SFR (e.g., \citealt{Madau98}), which is found to
have decreased by about one order of magnitude since its peak
(e.g.,\citealt{RosaGonzalez02} and references therein). Since very
strong absorbers are evolving away at redshifts coincident with the
decrease in SFR, they could be tracing the star formation sites of
galaxies, as previously suggested by \citet{Guillemin97} and
\citet{Churchill99}. Low-mass galaxies present this peak of SFR at
even lower redshifts (\citealt{Kauffmann04}).  In a study of $z <$~0.8
{\MgII} absorption of SDSS DR3, \citet{Bouche06} found an
anti-correlation between halo mass of galaxies and the equivalent
width of {\MgII} absorption present. Since the very strong {\MgII}
systems may show saturation in most of their profiles, the
equivalent width is determined primarily by the velocity dispersion of
the absorbing clouds. The required velocity spreads are consistent
with a starburst picture for the strongest {\MgII} systems, but not
with structure within individual virialized halos. \citet{Prochter06}
supported the claim that very strong {\MgII} absorbing structures are
related to superwinds rather than accreting gas in galaxy halos, in a
study over a larger redshift range ($0.35 < z < 2.7$) and based on the
kinematics of $W_r(\MgII)>1$~{\AA} absorbers. High resolution studies
of individual absorbers confirm that they show particular signatures
in their profiles that might be consistent with superwinds
(\citealt{Bond01}; \citealt{Ellison03}). Moreover, field imaging of a
subset of the strongest {\MgII} absorbers ($W_r(\MgII)>2.7$~{\AA}) at
low redshift ($0.42 < z < 0.84$) indicates that interactions, pairs,
and starburst related phenomena are likely to be present
(\citealt{Nestor07}).

Weak {\MgII} absorption systems have been found to evolve as
well. \citet{Narayanan05} and \citet{Narayanan07} found a peak at $z
=$~1.2 in the number per unit redshift, $dN/dz$, of {\MgII} absorbers
with $W_r(\MgII) <$~0.3 \AA.  Moreover, \citet{Narayanan08} find a
trend in the ratio of {\FeII} to {\MgII}: at higher redshift
($z>$~1.2), weak {\MgII} absorption systems with large values of
$W_r(\FeII\lambda2383)/W_r(\MgII)$ are rare. They suggested that this
trend could either be caused by an absence of high density, low
ionization gas at high-$z$ or the absence of enrichment by Type Ia
supernovae in weak {\MgII} clouds at high-$z$. They suggest that the
relatively few weak {\MgII} absorbers that are observed at high $z$
are from young stellar populations and thus $\alpha$--enhanced.

This paper investigates the evolution of very strong {\MgII}
absorption systems with redshift, particularly their {\FeII} to
{\MgII} ratios.  We enlarge the sample of strong {\MgII} absorption
presented by \citet{Mshar07}, and compare it to the analysis previously
carried out in \citet{Narayanan07} for weak {\MgII} absorbers. The
data and methods, as well as the systems found, are presented in
section \S\ref{sec:2}. Results involving evolution of the profiles and
the {\FeII}/{\MgII} ratio are shown in section \S\ref{sec:3}. We
summarize the results and discuss their implications in section
\S\ref{sec:4}.

\section{Data and Survey Methods}
\label{sec:2}

We analyzed 81 high resolution $R$~$\sim$~45,000 ($=6.7$~{\kms})
VLT/UVES quasar spectra retrieved from the ESO archive. This
particular set of spectra was first compiled for the
\citet{Narayanan07} search for weak {\MgIIdbl} absorbers, although the
spectra were originally obtained to facilitate a heterogenous range of
studies of the Ly$\alpha$ forest and of strong metal-line absorbers.
Because some lines of sight were observed because of previously known
strong absorbers, we expect a bias toward strong {\MgII} absorption.
Since our focus is on the evolution of absorber properties, and not on
tabulating a statistical distribution function of equivalent widths,
this bias does not affect our study.  However, we remain
alert to the fact that a particular kind of strong absorber could be
favored in this study if commonly observed by the original studies.
The list of quasars used, with information on each quasar observation,
can be found in Table 1 of \citet{Narayanan07}. The data reduction is
described in \citealt{Narayanan07}, section 2.1. The average
signal-to-noise ratio of the spectra is large ($>$ 30 pixel$^{-1}$
over the full wavelength coverage), which facilitates detection of
weak absorption components.

Although UVES spectra have a large wavelength coverage (3000 \AA\, to
1 $\mu$m), the choice of cross-disperser settings may cause breaks in
this coverage. We searched the same redshift path length available in
each quasar spectrum as did \citet{Narayanan07} (illustrated in their
Figure 1), which excludes the Ly\a forest and wavelength regions
contaminated by telluric features. We also exclude systems within
3,000 \kms of the quasar redshift in order to avoid contamination with
intrinsic associated absorption lines. 
Although strong systems are easier to spot
than weak systems in contaminated regions, we avoided such regions
because we could easily miss weaker subsystems associated with such
systems, which would bias our results\footnote{We define 'subsystem',
following the CV01 definition, as 'absorption features that are
separated by more than 3 pixels (i.e., a resolution element) of
continuum flux'. \MgII\, doublets found within 1,000 \kms\, of
each other were classified as subsystems that are part of the
same absorbing system.}. In fact, strong absorption systems are almost always
composed of more than one subsystem. 

We searched for strong {\MgII} ($W_r>$~0.3 \AA) systems along the 81 quasar 
lines of sight and found a total of 87 systems along 47 of the sight lines. 
We used a 5{\s} search criterion for {\MgII} \lb2796 and at
least 2.5{\s} for {\MgII} \lb2803, which was sufficient for
strong {\MgII} absorbers, especially due to the high
S/N of the spectra. After confirming the presence of {\MgIIdbl} by
visual inspection of the profile shapes, we looked for other ions that could
be present at the same redshift: {\FeII} \lb2374, \lb2383, \lb2587 and
\lb2600, and {\MgI} \lb2853. The {\FeII} and {\MgI} transitions can be
used for a better understanding of the kinematics of systems that
appeared saturated in {\MgII}. Since these ions are weaker than {\MgII}
we searched for them using a 3{\s} detection limit.
We also compared the profile shapes of the {\FeII} transitions and of
{\MgI} to those of {\MgIIdbl}, in order to identify possible blends
contaminating the profiles. We placed upper limits on equivalent widths
whenever such blends occurred.
Equivalent widths of all detected transitions were computed by a pixel
by pixel integration of the profiles, including all subsystems, as in
CV01.  The absorption redshift of each system was formally defined by
the optical-depth-weighted median of the {\MgII} \lb2796 profile,
following Appendix A in CV01.

\begin{figure}
\begin{center}
\includegraphics[width=7cm]{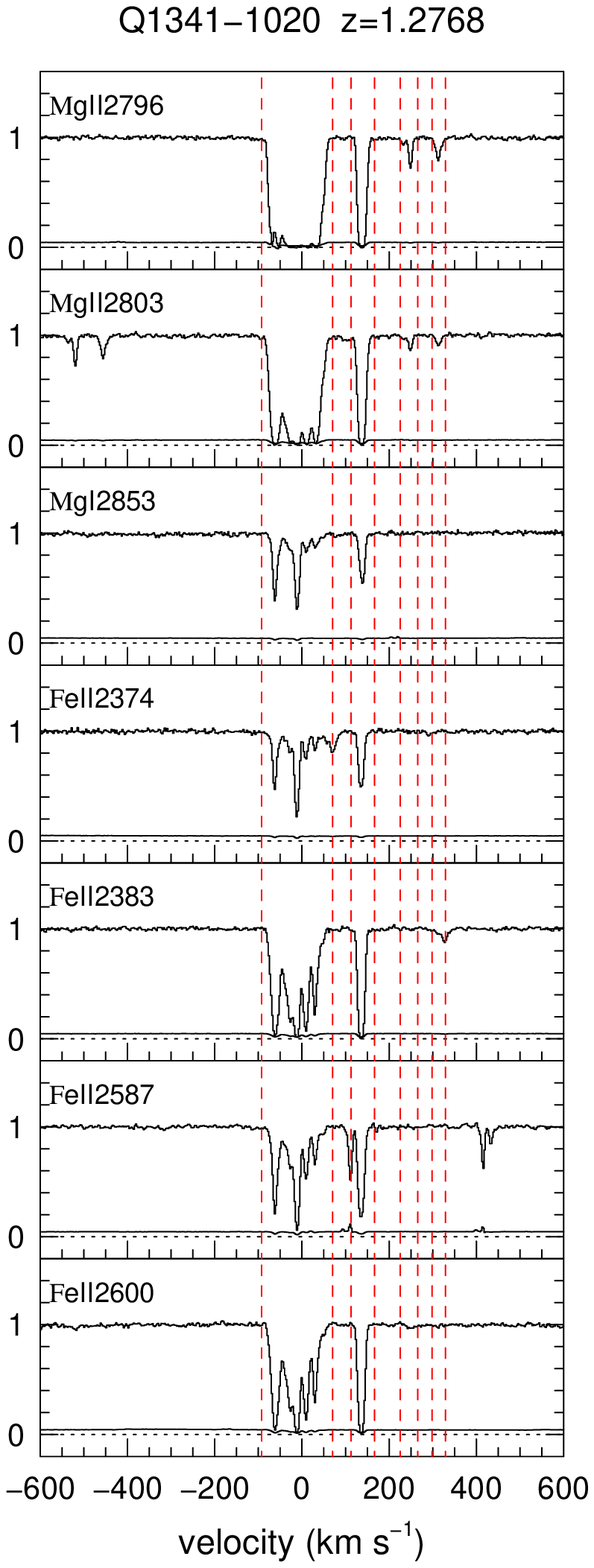}
\caption{VLT/UVES profiles of {\MgII} \lb2796, {\MgII} \lb2803, {\MgI} \lb2853,
{\FeII} \lb2374, {\FeII} \lb2383, {\FeII} \lb2587, and
{\FeII} \lb2600 (if covered) for the strong {\MgII}
absorption lines systems in our sample. The spectra are
normalized and the different transitions are aligned in
velocity space (zero point corresponds to the optical depth
mean of the {\MgII} \lb2796 profile). The vertical dashed
lines delimit the regions in which {\MgII} \lb2796 is detected. {\it [See the
electronic edition of the Journal for Figs 1.2 to
1.87.]}}
\label{profiles}
\end{center}
\end{figure}

The 87 strong {\MgII} systems are shown in Figure Set \ref{profiles} and
listed in Table \ref{data}.
Figure Set \ref{profiles} includes profiles of {\MgIIdbl}, and of {\MgI}
\lb2853, {\FeII} \lb2374, {\FeII} \lb2383, {\FeII} \lb2587 and {\FeII}
\lb2600, whenever the spectral region was covered. In these
figures, the spectra are normalized and the different transitions are
aligned in velocity space, centered at the median of the apparent optical
depth of the {\MgII}~\lb2796 profile. We included vertical dashed lines to
delineate the separate subsystems, although our equivalent width
measurements are only tabulated for the systems as a whole.
 
\begin{figure}
\begin{center}
\includegraphics[width=8cm]{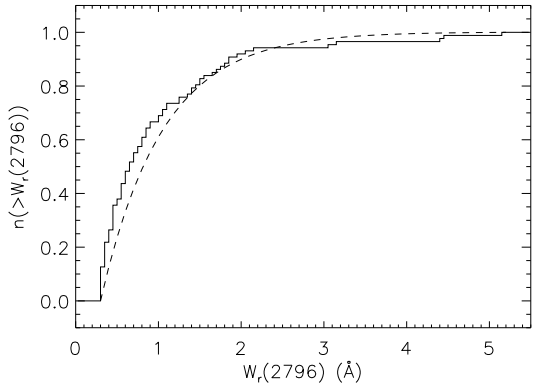}
\caption{Cumulative {\MgII} \lb2796 equivalent width distribution for the systems detected
in this study (shown as a solid line). Dashed line represents the distribution of systems in the larger sample of \citet{Nestor05} using the median of our sample ($\langle z \rangle$= 1.268).}
\label{cumdist}
\end{center}
\end{figure}

\begin{figure}
\begin{center}
\includegraphics[width=8cm]{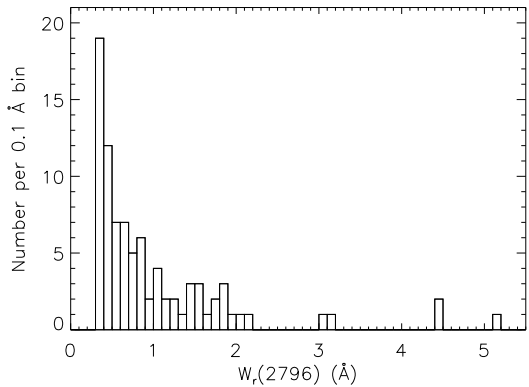}
\caption{Distribution of the rest-frame equivalent width ($W_{r}$) of {\MgII} \lb2796
  for the systems detected in this
  study. Most of the systems found lie in the lower part of the $W_r$
  range. The very strong systems comprise $\sim$33\% (29/87) of the
  total sample. Only $\sim$~8\% (7/87) of all systems present
  $W_r \geq$~2~{\AA}.}
\label{Wrdist}
\end{center}
\end{figure}

Table \ref{data} lists the rest-frame equivalent widths or equivalent
width limits for the {\MgII}, {\FeII}, and {\MgI} transitions of the
87 strong systems.  Among them, 58 systems are moderately strong (0.3
\AA~$<W_r(\MgII) <$~1 \AA) and 29 systems are very strong absorbers
($W_r(\MgII) >$~ 1 \AA).  Blanks in Table \ref{data} represent cases
where the wavelength of the relevant transition was not covered in the
VLT/UVES spectrum. In cases where the {\FeII} transition or {\MgI} was
covered, but not detected, 3\s upper limits are given.  We visually
inspected every {\FeII} absorption feature, and carefully compared
their profiles to each other and to the {\MgII} profiles, in order to
confirm that they were consistent.  We rejected deviant parts of these
components, mostly caused by blends with different transitions from
systems at other redshifts.  In cases where blends prevented the
measurement of an accurate equivalent width, we place lower limits by
avoiding the blended region and upper limits by including it. These
results appear as a range of values in Table \ref{data} (e.g.,
0.354--0.426~{\AA} for the $W_r(2587)$ measurement for the absorption
system at $z_{abs}$~=1.672 in the spectrum of Q0237-0023). When
systems suffered from severe blending such that the above mentioned
procedure was not possible, we just include $W_r$ upper
limits. {\FeII} \lb2383 and {\FeII} \lb2600 are the strongest of the
{\FeII} transitions, and thus when {\MgII} is saturated they may also
be saturated. Therefore, we measured other weaker {\FeII} transitions
({\FeII} \lb2374 and {\FeII} \lb2587), that are less likely to be
saturated, and thus provide more leverage on an accurate measurement
of the {\FeII}.

\begin{figure}
\begin{center}
\includegraphics[width=8cm]{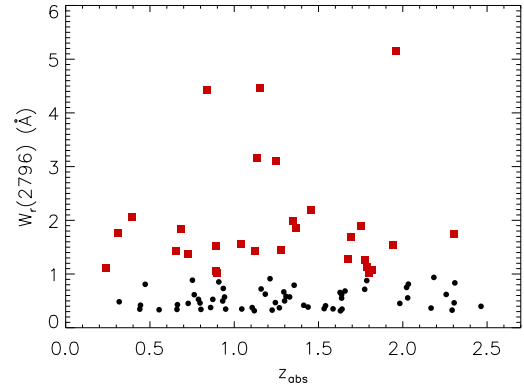}
\includegraphics[width=8cm]{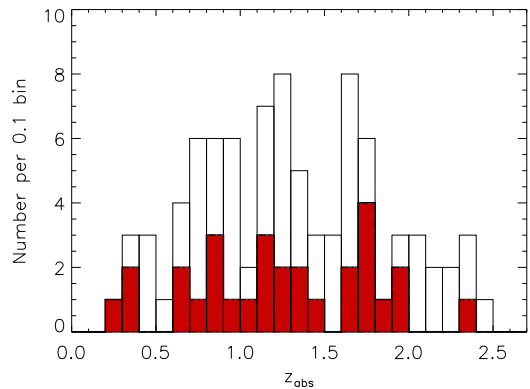}
\caption{Redshift distribution of the strong {\MgII} \lb2796 in this
study. The top panel shows rest-frame equivalent width of {\MgII}
versus the absorption redshift. Black circles represent the moderately strong absorption systems
(0.3~$< W_r <$~1.0~{\AA}) and red squares are used for very
strong systems ($W_r>$~1.0~{\AA}). Errors in $W_{r}$ are smaller than the symbol sizes. 
The bottom panel shows a histogram of
the redshift distribution. The total histogram represents the distribution of
all systems (all symbols in top panel), and the
red area the distribution of very strong systems with $W_r(\MgII)>$~1.0~{\AA} (red squares in top panel).}
\label{zadist}
\end{center}
\end{figure}

In Figures \ref{cumdist} to \ref{zadist} we characterize the {\MgII}
absorption properties of our sample. Figure \ref{cumdist} shows the
cumulative equivalent width distribution of {\MgII} \lb2796 for our sample
of strong {\MgII} absorbers.  Figure \ref{Wrdist} shows the binned
rest-frame {\MgII} \lb2796 equivalent width distribution.
The distribution of equivalent widths is similar to that for the
unbiased survey of \citet{Nestor05}, however our sample has a slightly
larger number of moderately strong systems relative to very strong systems.
This small difference does not represent a problem for the present study, since our aim is
to compare the same types of systems at different redshifts.

Figure \ref{zadist} presents the distribution of the {\MgII}
absorption systems in equivalent width-redshift ($W_r$--$z_{abs}$)
parameter space. The top panel shows $W_r(\MgII)$ versus redshift for
the 87 strong {\MgII} absorption systems in our study.  Throughout
this paper we display moderately strong {\MgII} absorption systems
(0.3~$< W_r <$~1.0~{\AA}) as black circles and very strong {\MgII}
systems ($W_r>$~1.0~{\AA}) as red squares.  The bottom panel shows the
binned absorption redshift ($z_{abs}$) distribution of all the systems
in top panel (in white). The 87 systems are roughly uniformly
distributed over the absorption redshift range, from $z_{abs}=$~0.238
to 2.464, with their mean at $\langle z \rangle=$~1.297. The redshift
histogram for only the very strong {\MgII} systems (red/shaded) is
superimposed.  The distributions in redshift for moderately strong
($\langle z\rangle$=1.322) and very strong systems ($\langle
z\rangle$=1.247) are not different at a statistically significant level
(the K--S tests shows $P$(K--S)=0.83 that they are drawn from the same
distribution). However, we note that there is only one very strong
system with $z>2$.


\section{Results}
\label{sec:3}

\begin{figure*}
\begin{minipage}{180mm}
\begin{center}
\includegraphics[height=6.3cm]{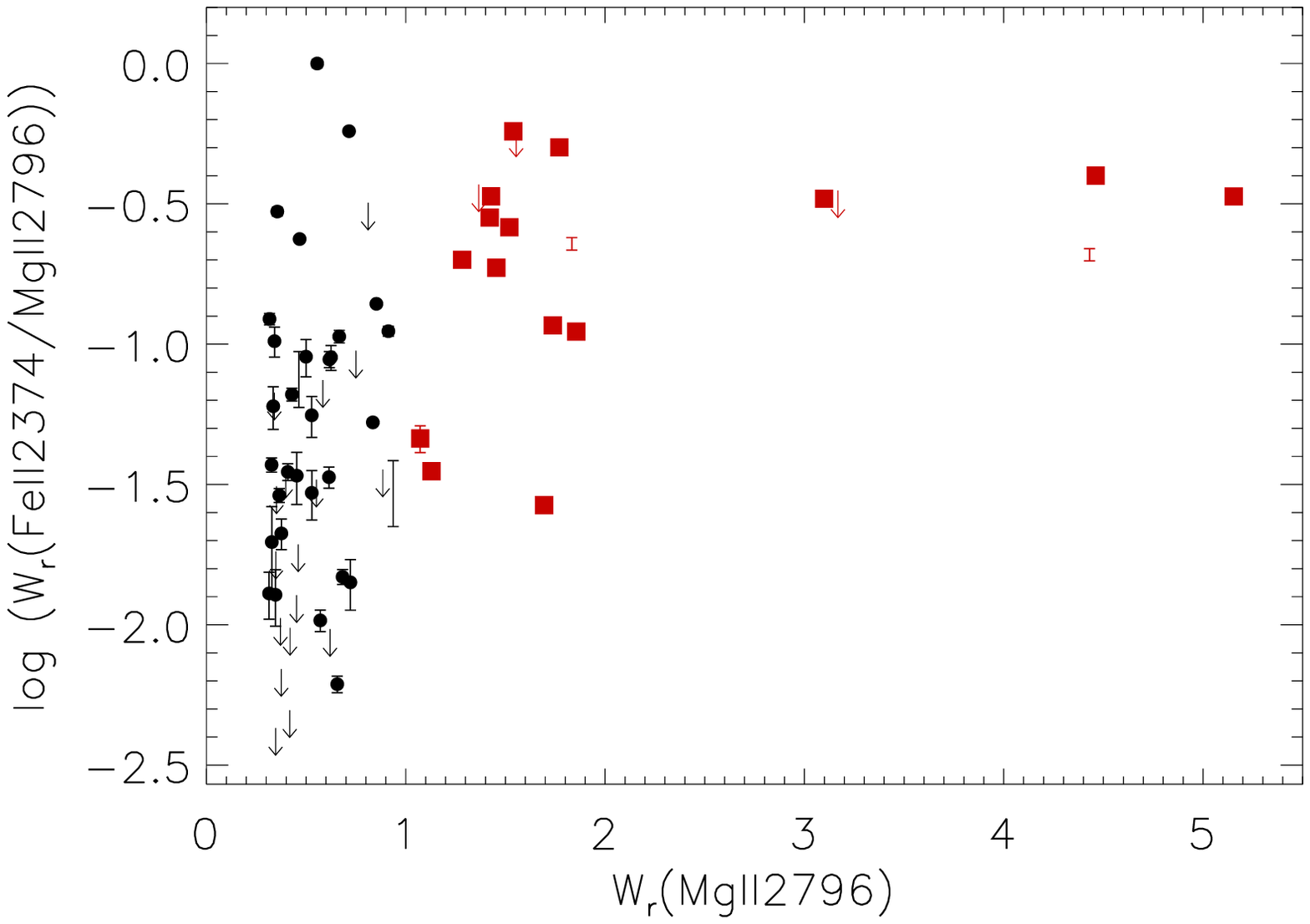}
\includegraphics[height=6.3cm]{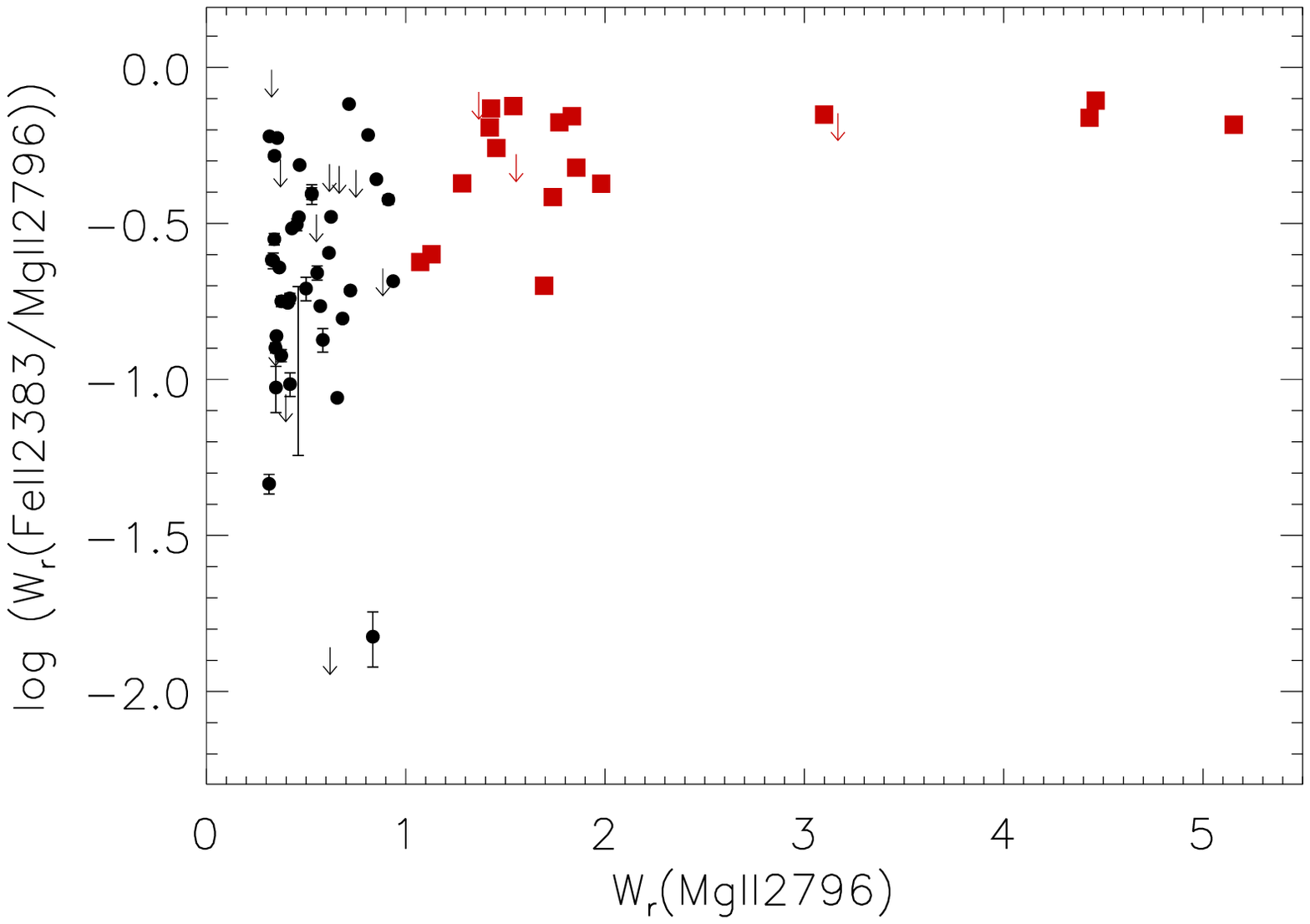}
\includegraphics[height=6.3cm]{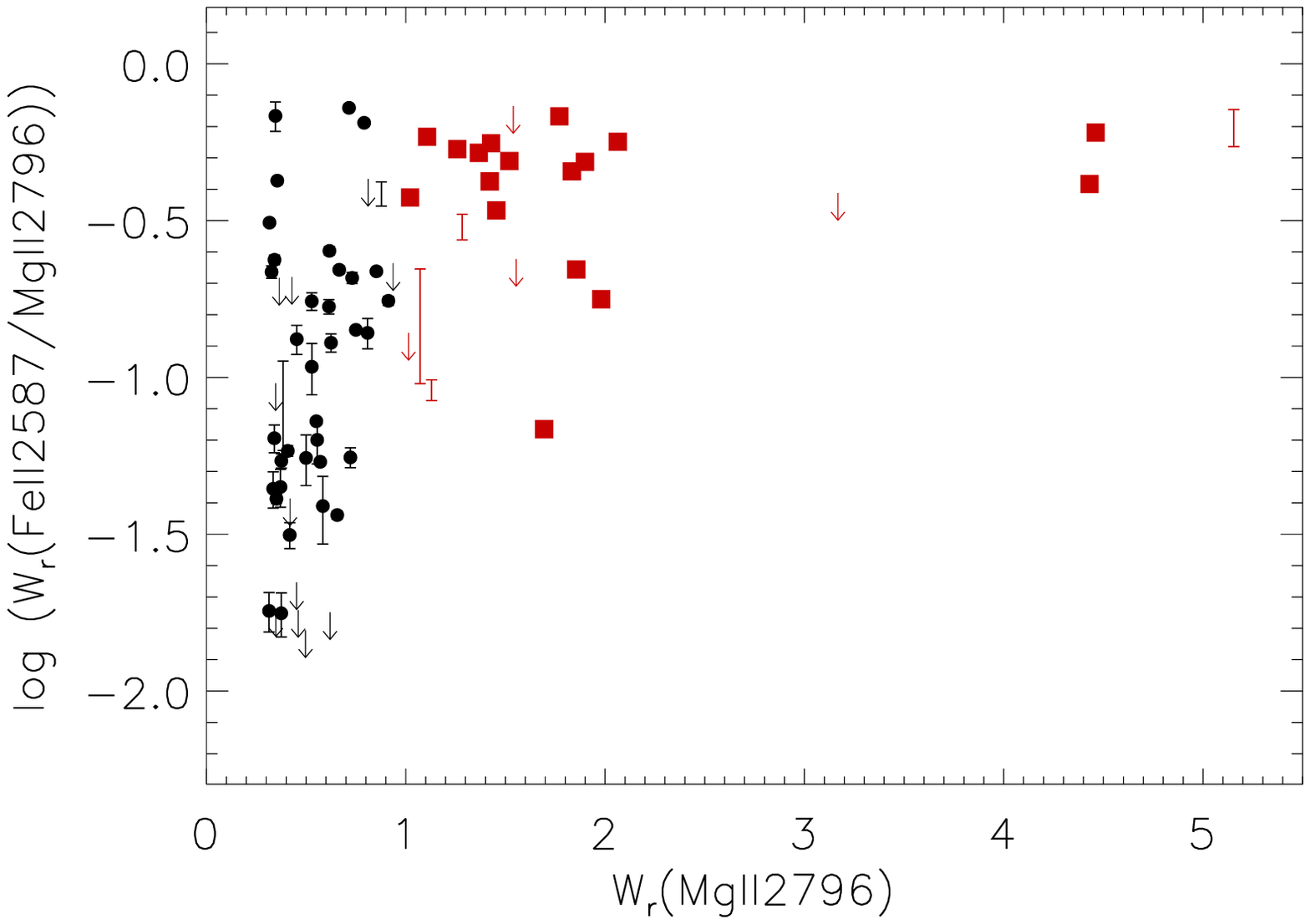}
\includegraphics[height=6.3cm]{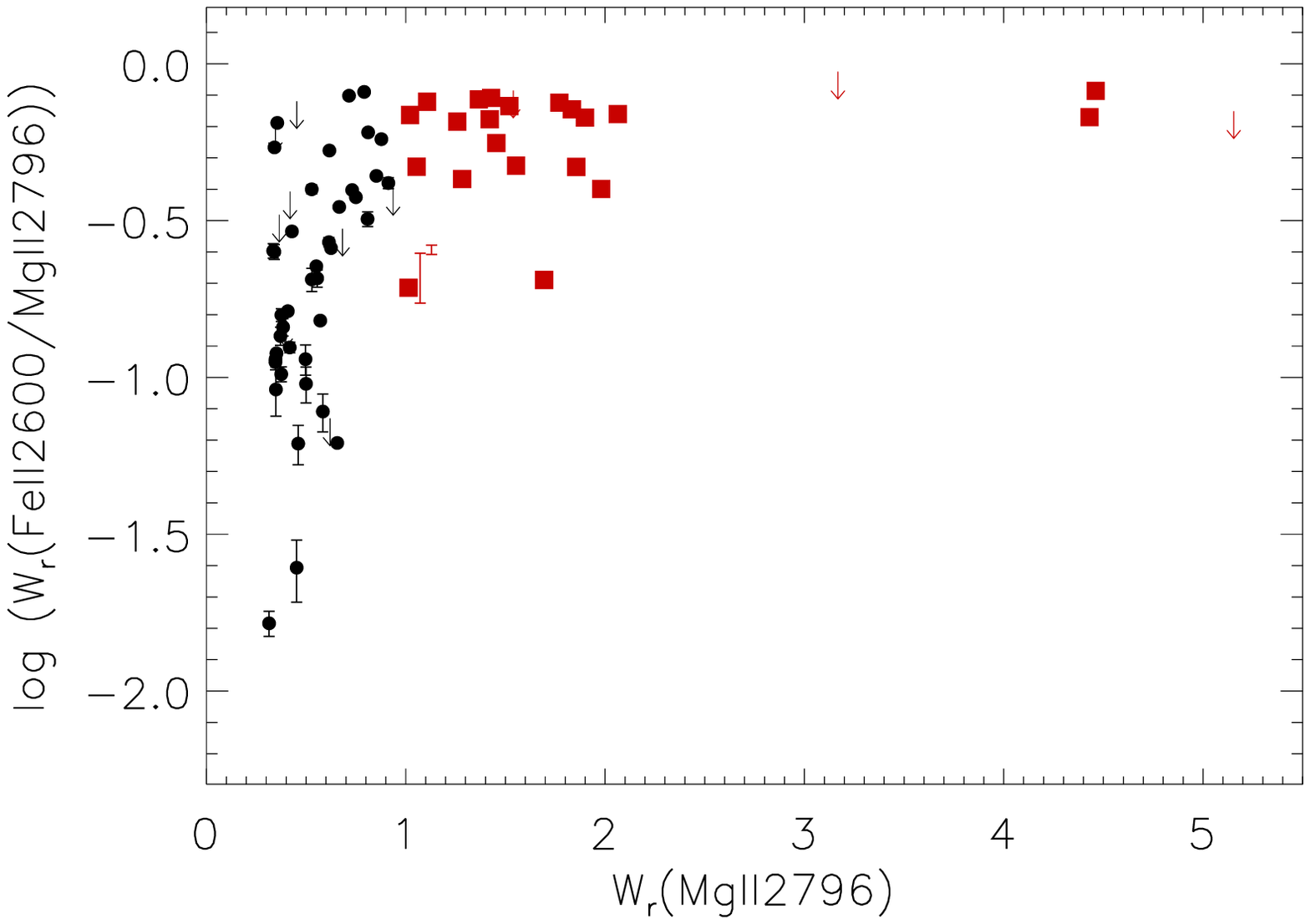}
\caption{Ratio of $W_r(\FeII)$ to $W_r(\MgII)$, for the four {\FeII}
  transitions measured in this study, versus equivalent width of the
  blue member of the {\MgII} doublet ($W_r(\MgII)$). Black circles
  represent moderately strong {\MgII} absorption (0.3~$< W_r <$~1.0~{\AA}) and red squares very
  strong {\MgII} absorption ($W_r>$~1.0~{\AA}). Arrows indicate upper limits (for blends
  in {\FeII} as well as for cases with no detection), and
  error bars with no central symbol indicate cases where we
  estimated an upper and lower limit value for the {\FeII} measurement
  (see \S\ref{sec:2}). The number of data points is not the same among the four panels due to the non-uniform coverage of the {\FeII} transitions in the VLT/UVES spectra. As the values of $W_r(\MgII)$ increases, $W_r(\FeII)$ values tend to increase as well, and when both profiles show signs of saturation, the ratio $W_r$/$W_r(\FeII)$ approaches unity. The figure shows how {\FeII} \lb2374 and {\FeII} \lb2587 display a larger range of $W_r$/$W_r(\FeII)$ values than {\FeII} \lb2383 and {\FeII} \lb2600, because the former two transitions are weaker than the latter.}
\label{ratioMgII}
\end{center}
\end{minipage}
\end{figure*}

\begin{figure*}
\begin{minipage}{180mm}
\begin{center}
\includegraphics[height=6.3cm]{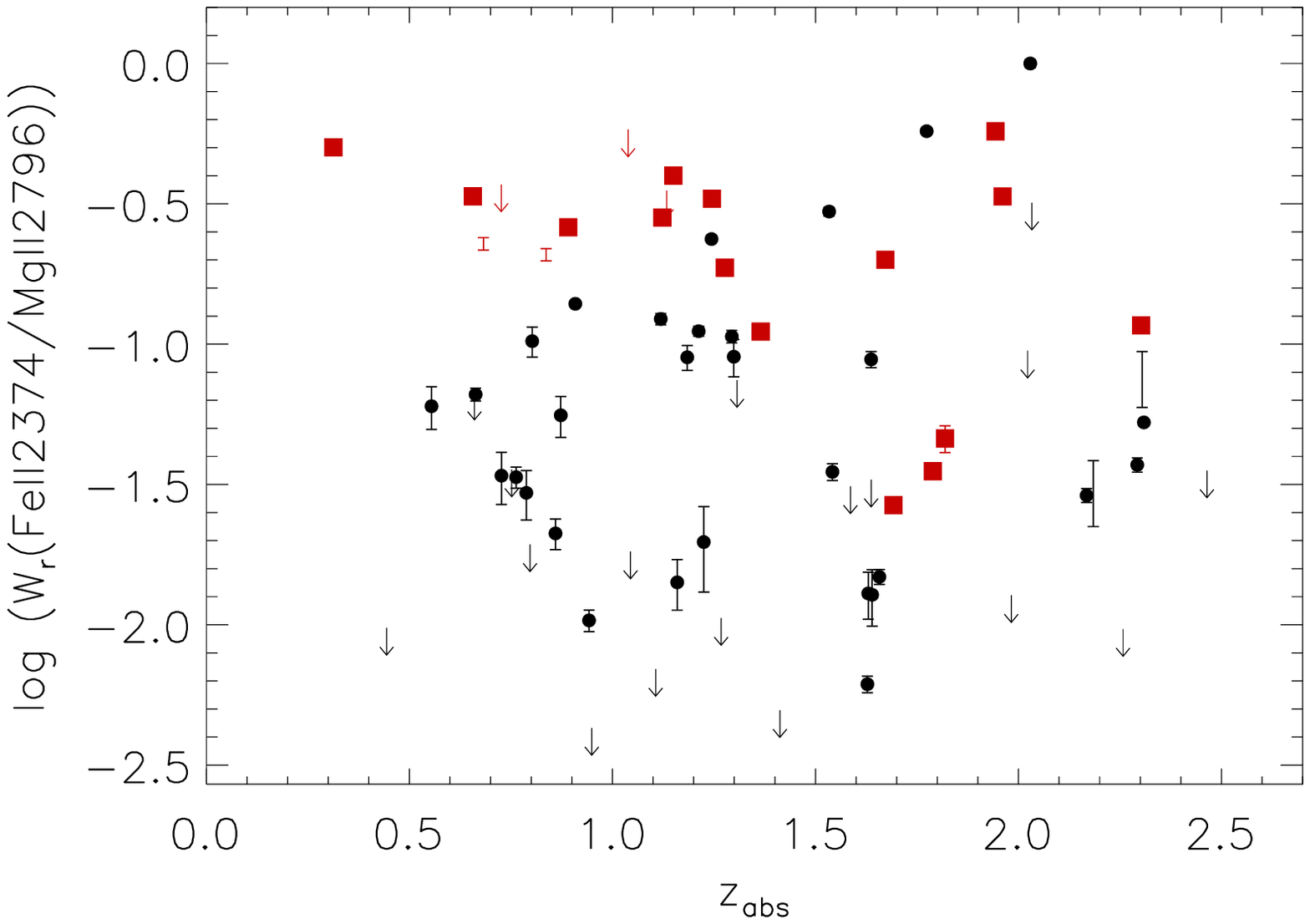}
\includegraphics[height=6.3cm]{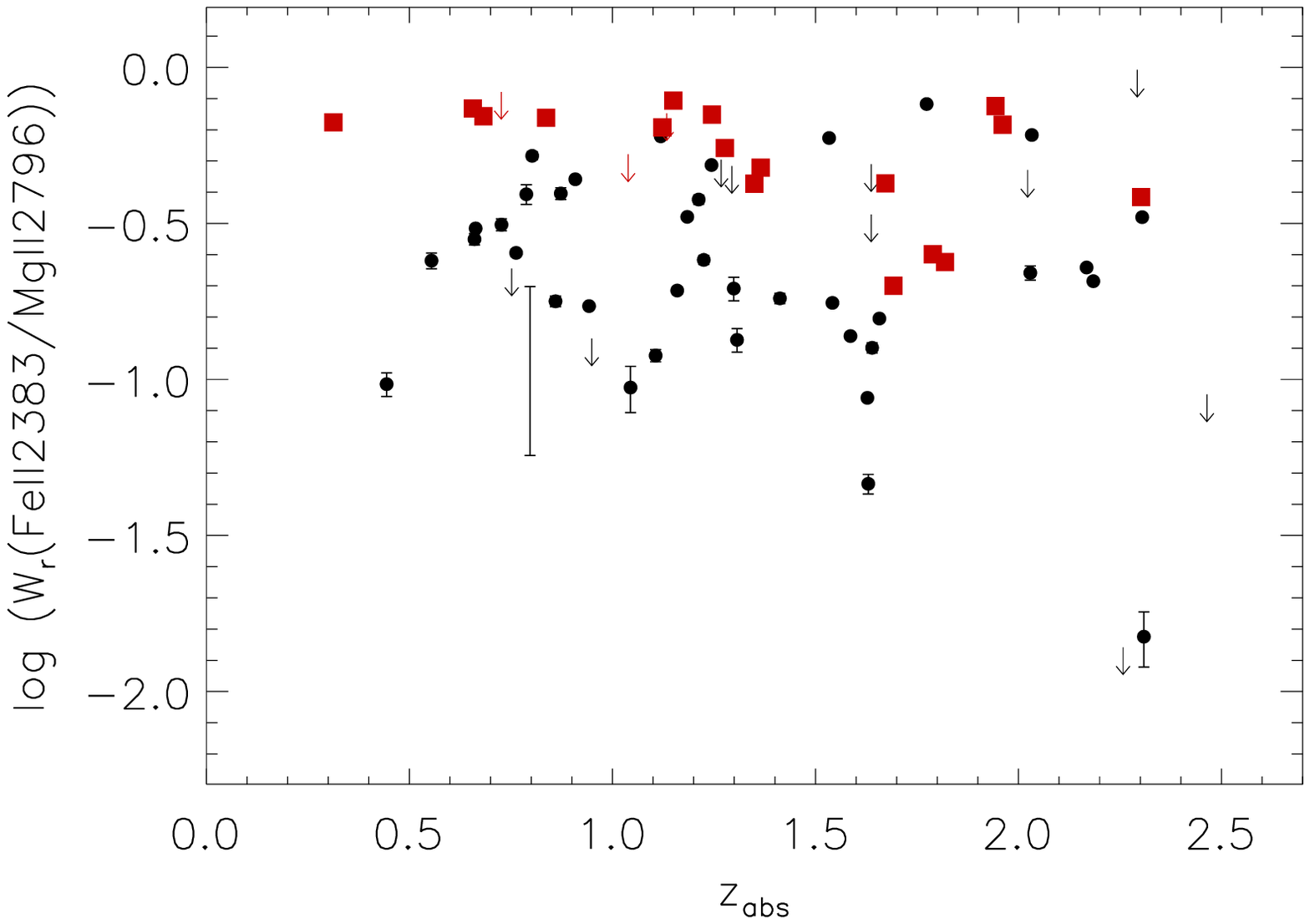}
\includegraphics[height=6.3cm]{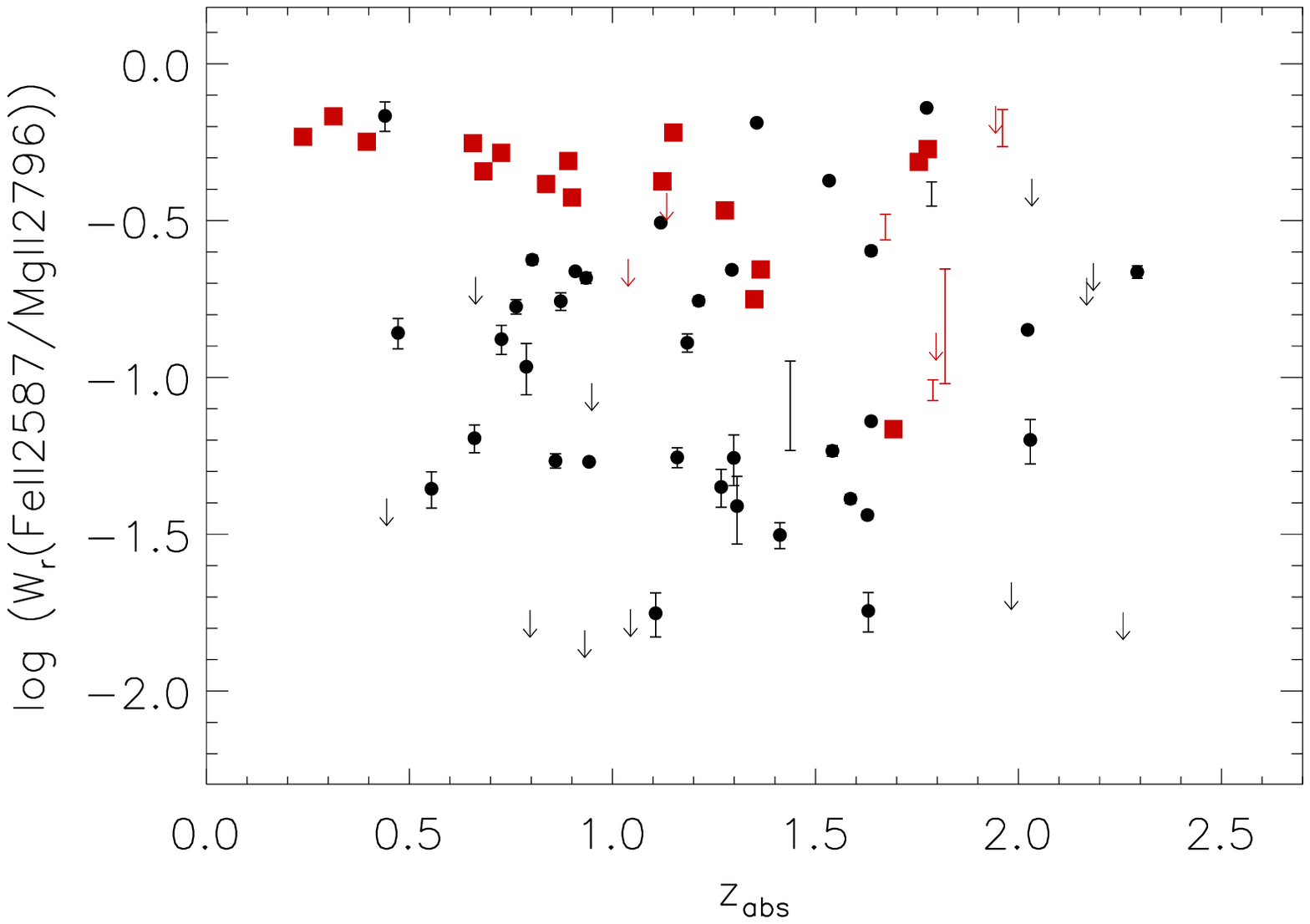}
\includegraphics[height=6.3cm]{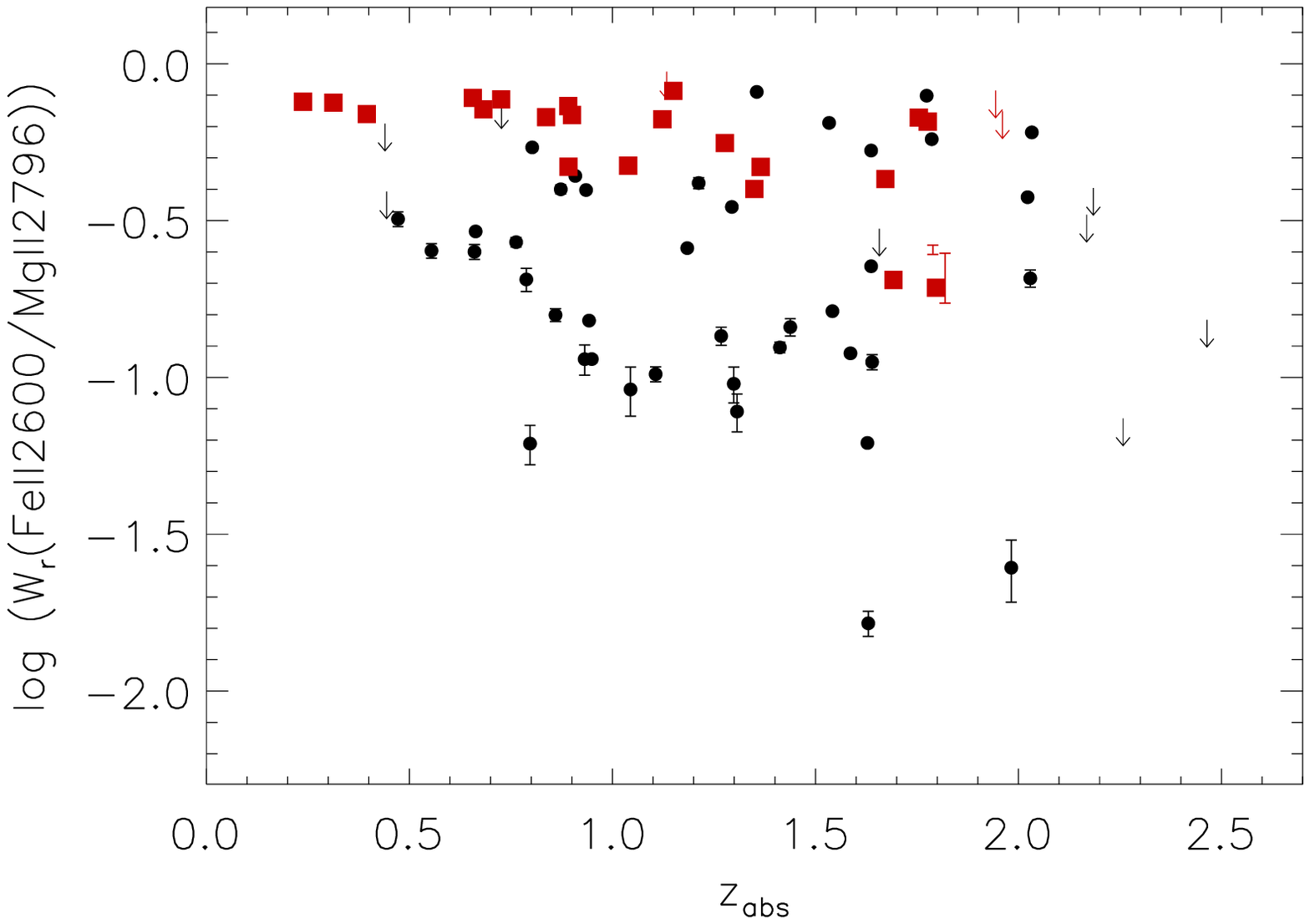}
\caption{Ratio of $W_r(\FeII)/W_r(\MgII)$ versus absorption redshift ($z_{abs}$) for the four {\FeII}
transitions studied. Symbols are equivalent to those in Figure \ref{ratioMgII}. The distribution of moderately strong {\MgII} absorbers seems to cover a large range of $W_r(\FeII)/W_r(\MgII)$ at all redshifts. On the contrary, very strong {\MgII} absorbers at lower redshifts ($z_{abs}<$~1.2) appear to cluster at high values of  $W_r(\FeII)/W_r(\MgII)$, which they do not do at higher redshifts.}
\label{ratioz}
\end{center}
\end{minipage}
\end{figure*}

As mentioned in \S\ref{sec:1}, the {\Fe}/{\Mg} abundance ratio is a
tracer of the star formation history of a galaxy. For typical physical
properties of interstellar gas, most of the {\Mg} and {\Fe} are in the
form of {\MgII} and {\FeII}, respectively. Thus, although we detected
{\MgI} ({\FeI} is almost never present, however, see \citealt{Jones10}),
hereafter we will focus on {\MgII} and {\FeII}. For some of the strong
absorbers in this study, and more likely for the very strong
absorbers, {\FeII} \lb2383 and {\FeII} \lb2600 are saturated, as
well as {\MgII} (see Figure set \ref{profiles}).  Thus for these
{\FeII} transitions, there is a slower $W_r(\FeII)$ increase, with
increasing {\FeII} column density, reducing the leverage of the
$W_r(\FeII)/W_r(\MgII)$ ratio in determining evolutionary trends.  We
compensate for this by using weaker {\FeII} transitions, such as
\lb2374 and \lb2587, which are less saturated, and for many systems do
not show evidence of saturation.  Since {\FeII} \lb2587 is more often
covered in this VLT/UVES sample, we often emphasize it in drawing
conclusions.

Figure \ref{ratioMgII} shows the distribution of ratios of
$W_r(\FeII)/W_r(\MgII)$ with respect to $W_r(\MgII)$ for the four
{\FeII} transitions measured in this study: {\FeII} \lb2374, {\FeII}
\lb2383, {\FeII} \lb2587, and {\FeII} \lb2600.  The ratio of
$W_r(\FeII)/W_r(\MgII)$ approaches unity as $W_r(\MgII)$ increases,
since both the {\FeII} and the {\MgII} are saturated for the strongest
systems.  {\FeII} \lb2374 and {\FeII} \lb2587 show the largest range
of ratio values for the very strong absorbers since they are less
saturated than the other two {\FeII} transitions.  Although our number
statistics are small for values of $W_r(\MgII)>$~2.5\AA, all of our
data points in that range have large ratios, fairly close to $1$, even
for {\FeII} \lb2374 and {\FeII} \lb2587.

Figure \ref{ratioz} shows the distribution of ratios of
$W_r(\FeII)/W_r(\MgII)$ in absorption redshift ($z_{abs}$), again for
the four {\FeII} ions measured in this study. We can see that these
distributions for the two different absorption classes (moderately
strong and very strong) show different trends.  While moderately
strong {\MgII} absorption systems (black circles) present a wide range
of $W_r(\FeII)/W_r(\MgII)$ ratios, independent of redshift, very
strong (red squares) {\MgII} absorbers are preferentially present in
certain regions of the $W_r$--$z_{abs}$ parameter space.  For very
strong {\MgII} absorbers (red symbols) at lower redshifts
($z_{abs}<$~1.2 in our sample, see below) $W_r(\FeII)/W_r(\MgII)$ is
concentrated at higher values, and there are few absorption systems with
small ratios of $W_r(\FeII)/W_r(\MgII)$. As redshift increases, a
wider range of the ratio is found.  As we showed in Figure
\ref{zadist}, there are a sufficient number of very strong systems at
low redshift (as compared to high redshift) that we can be confident
that this trend is not due to small number statistics.

The difference in the distribution of $W_r(\FeII)/W_r(\MgII)$ between
low and high redshift, for very strong absorbers, is clear from visual
inspection of Figure \ref{ratioz}: at low redshift, the very strong
{\MgII} absorbers show near-unity $W_r$ ratios while there is a much
larger spread at high redshift, extending to smaller $W_r$ ratio
values.  We proceed to quantify the significance of this trend by using several
statistical techniques (Table \ref{stats}). We investigate the
probability that two subsamples were derived from the same population
by performing Kolmogorov-Smirnov (K--S) and Anderson-Darling (A--D)
tests. The A--D is a modification of the K--S test and gives more
weight to differences in the tails of the distribution of the data
(\citealt{Scholz87}).
Table \ref{stats} includes the comparisons between the $W_r$ ratio
distributions of moderately strong and very strong systems, and
between high and low redshift subsamples of each. To
divide each sample into low-redshift and high-redshift subsamples we
compute the median of the $z_{abs}$ values, excluding upper limits,
for each {\FeII} transition and average them, which yields
$<z_{abs}> =$~1.20.

The results in Table \ref{stats} confirm that the
$W_r(\FeII)/W_r(\MgII)$ distribution for moderately strong absorption
systems is not consistent with that for the very strong absorption
system population.  We see that this difference is more significant in
the low redshift bin (see second row of Table \ref{stats}) with
$W_r(\FeII)/W_r(\MgII)$ approaching unity for the very strong systems
(see Fig.~\ref{ratioz}).  Through bootstrapping, we found that in only
$\leq 0.8\%$ of the realizations of the moderately strong sample at
low redshift were there as many large $W_r(\FeII)/W_r(\MgII)$ values
(larger than the median of $W_r(\FeII)/W_r(\MgII)$) as in the very
strong sample at low redshift (see Table \ref{stats} for more
details).  At higher redshifts, there is still some difference between
the $W_r(\FeII)/W_r(\MgII)$ values for very strong and moderately
strong samples (see third row of Table \ref{stats}), but it is not as
pronounced.

A comparison of the low and high redshift subsamples of very strong
absorption systems shows that there is only a small probability that
they share the same parent population (see fourth row of Table
\ref{stats}).  The largest probability of $W_r(\FeII)/W_r(\MgII)$ of
the low and high redshift subsamples coming from the same parent
population is for {\FeII} \lb2374, $P$(K--S)=0.04.  Moderately strong
absorption systems show no such difference (see fifth row of Table
\ref{stats}).  Thus we conclude that there is a significant evolution
in redshift in the $W_r(\FeII)/W_r(\MgII)$ distribution for very strong {\MgII}
absorbers, but not for moderately strong absorbers.

The left-bottom quadrant of each of the Figure \ref{ratioz} panels is
almost, but not completely, devoid of very strong systems (red squares), thus we
would like to comment on the exceptions that do lie in this quadrant.
Also, there are a few upper
limits on the {\FeII} measurements of each transition (red arrows), due to
significant contamination, and some of these values could, in
principle, come down into this quadrant if they could be accurately
measured.  In the case of the system at $z_{abs}=$~0.8919 towards
Q0300+0048 the spectrum is
very noisy ($S/N < 5$; see Figure \ref{profiles}) around the {\FeII}
\lb2374, {\FeII} \lb2383, and {\FeII} \lb2587 transitions, so a point
does not appear on those panels. However, the quality of the spectrum
at the position of {\FeII} \lb2600 is higher, and it does appear to
yield a relatively low value of $W_r(2600)/W_r(\MgII)$, which
is apparent in the bottom-right panel of Figure \ref{ratioz}.  A different
situation applies in the case of the $z_{abs}=$~1.0387 system towards CTQ0298,
another of the low redshift data points with a low $W_r(\FeII)/W_r(\MgII)$
value in the bottom-right panel. The {\FeII} absorption for the transitions shortward of 2600~{\AA} lie in the
Ly$\alpha$ forest of CTQ0298, and the resulting blends only allow us
to place upper limits on $W_r(\FeII)$ for those transitions.  In the
case of the $z_{abs} =$~0.7261 system toward Q0453-0423, {\FeII} \lb2374
and {\FeII} \lb2383 appear to be also blended with other ions, but it is likely
that their actual $W_r$ values are close to the upper
limits, which implies that their $W_r(\FeII)/W_r(\MgII)$ values are
truly in the upper-left quadrant. We have good estimates for the unblended {\FeII} \lb2587 and 
{\FeII} \lb2600 measurements which show large values of the $W_r(\FeII)/W_r(\MgII)$.
Finally, all of the {\FeII}
transitions associated with the $z_{abs}=$~1.1335 system toward
Q1621-0042 suffer from severe blending in the Ly$\alpha$ forest, thus
it is impossible to determine the true $W_r$ values.  Having
considered these points, we conclude that, with a few possible exceptions, there
is a true deficit of small $W_r(\FeII)/W_r(\MgII)$ values for very
strong {\MgII} absorbers at low redshift.

\subsection{{\MgII} profile evolution}
\label{3.1}

\begin{figure*}
 \begin{minipage}{180mm}
\begin{center}
\vspace{1cm}
\includegraphics[width=15cm]{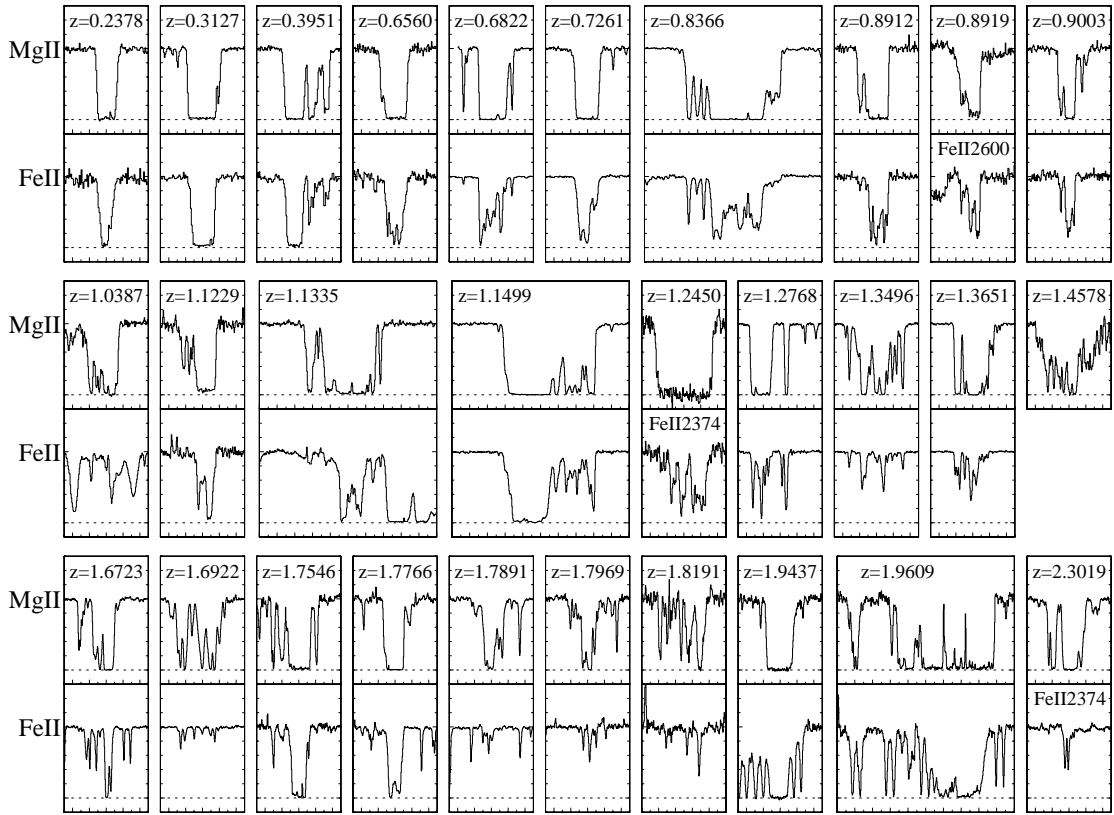}
\caption{Normalized {\MgII} and {\FeII} profiles of the very strong {\MgII} systems in our sample, in increasing order of $z_{abs}$, which is indicated at the top of each spectrum. {\MgII} \lb2796 and {\FeII} \lb2587 are plotted unless otherwise noted. There is no spectral coverage of {\FeII} in the spectrum of Q1418-064 for the absorber at $z_{abs}=$1.4578. Velocity ranges are (-250,250) {\kms}  in the single boxes and (-500,500) {\kms} in the double boxes, thus the velocity scale is the same in all boxes, except for the case of Q0551-3637 at $z_{abs}=$~1.9609, where the range is (-600,400) in order to include the entire profile. Dotted horizontal lines indicate flux zero.}
\label{mosaic}
\end{center}
 \end{minipage}
\end{figure*}

Figure \ref{ratioz} shows there is an evolution of the
$W_r(\FeII)/W_r(\MgII)$ ratio of very strong {\MgII} absorbers. We
investigate the profile shape evolution of {\MgII} for the these
systems in order to consider how it might affect the
$W_r(\FeII)/W_r(\MgII)$ ratio.

Figure \ref{mosaic} includes all the normalized profiles of very
strong {\MgII} systems in our sample, and their corresponding {\FeII} \lb2587,
ordered by increasing $z_{abs}$. In cases where {\FeII} \lb2587 is not covered,
another transition is used as indicated on the panels in the figure.
We find that the profile shape of {\MgII} varies from a predominance of
``boxy'' profiles at low redshift, to the inclusion of more
kinematically extended and less saturated {\MgII} profiles at high
redshift, as would be expected for outflows/superwinds.

\begin{figure*}
 \begin{minipage}{180mm}
 \vspace{-2cm}
\begin{center}
\vspace{1.6cm}
\includegraphics[width=20.8cm,angle=90]{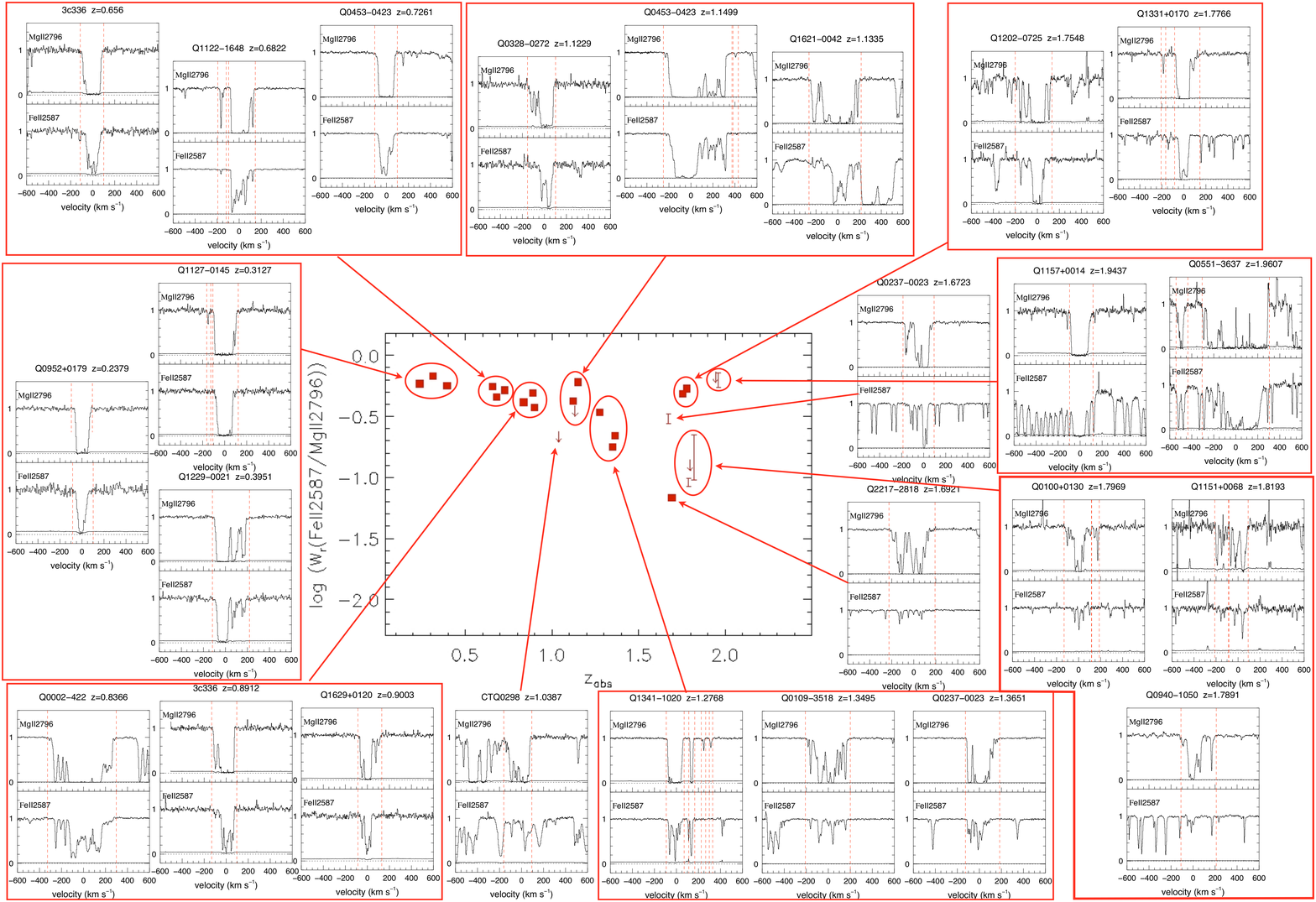}
\caption{{\MgII} \lb2796 and {\FeII} \lb2587 profiles of very strong {\MgII} systems ($W_r>$~1.0~{\AA}) in the parameter space of $\log(W_r(\FeII)/W_r(\MgII))$ vs. $z$, which was previously displayed in Fig.\ref{ratioz} (symbols are equivalent). For ease of location, we have grouped contiguous systems in boxes and circles. The figure shows that while ``boxy'' profiles appear at both low and high redshift, they are more predominant at low redshift. On the other hand, systems with ``outflow/superwind'' type profiles (less saturated and showing larger velocity spreads) are only present at high redshift, and correspond to the systems with smaller values of $W_r(\FeII)/W_r(\MgII)$.}
\label{Fig13}
\end{center}
 \end{minipage}
\end{figure*}

Figure \ref{Fig13} places the very strong {\MgII} and {\FeII} \lb2587
absorption profiles at their corresponding points of the
$W_r(\FeII)/W_r(\MgII)$--$z_{abs}$ parameter space. This allows us to
examine the relationship between profile shape and the
$W_r(\FeII)/W_r(\MgII)$ ratio. We can see from Figure \ref{Fig13} that
the evolution in the {\MgII} profile kinematics with redshift seen in
Figure \ref{mosaic} is related to the evolution seen in Figure
\ref{ratioz} of the values of $W_r(\FeII)/W_r(\MgII)$ of the very
strong {\MgII} systems. Most of the kinematically complex, broad
profiles correspond to small $W_r(\FeII)/W_r(\MgII)$ ratios at high
redshift. We discuss this effect in Sec \S\ref{sec4.1} and, in Sec \S\ref{sec4.2},
contrast it with the previous analysis of $W_r(\FeII)/W_r(\MgII)$ for
weak {\MgII} systems.

\subsection{Kinematical properties}
\label{3.2}

\begin{figure}
\begin{center}
\includegraphics[width=8cm]{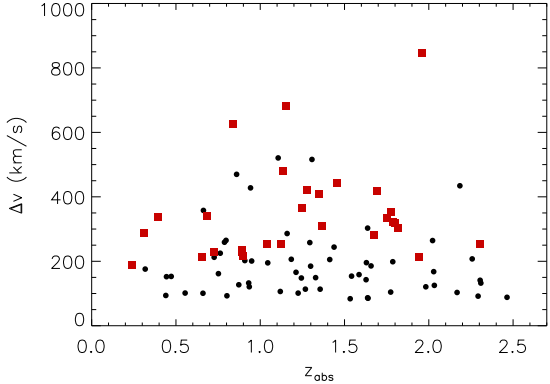}
\caption{Velocity spread versus absorption redshift of the 87 systems analyzed. Symbols are equivalent to those in Figure \ref{ratioMgII}. Strong {\MgII} absorption profiles (black circles) show a homogeneous distribution with redshift, but very strong {\MgII} absorption profiles (red squares) show a significant difference in their median value of velocity spread between low ($z_{abs}<$~1.2) and high ($z_{abs}>$~1.2) redshift.}
\label{deltav}
\end{center}
\end{figure}

To quantify the evolution we describe in section \ref{3.1}, we also
perform kinematic analysis on the {\MgII} profiles of the systems in
our sample. We saw in Figures \ref{mosaic} and \ref{Fig13} that systems
with larger kinematically spreads tend to be more predominant at higher
redshifts. We quantify this effect by calculating the velocity spread,
which is a measure of the full absorption line width (in units of
velocity), defined as

\begin{displaymath}
\Delta v = v_{low} - v_{high},
\end{displaymath}
\noindent where $v_{low}$ and $v_{high}$ correspond to the minimum and
maximum velocity of the absorption, defined by a 3$\sigma$ absorption
detection.  Figure \ref{deltav} shows the velocity spreads ($\Delta
v$) of the {\MgII} \lb2796 profiles in our sample. The distribution of
$\Delta v$ for moderately strong systems (black circles) is similar at
all redshifts, but we find differences between the very strong
absorbers (red squares) at low and high redshift. The median of the
low and high redshift values of $\Delta v$ differs (255 and 334 \kms,
respectively) and the K--S test shows there is a probability of only
0.03 that both subsamples derive from the same population.  If we omit
the outliers of $\Delta v > 600$ km s$^{-1}$, the K--S test shows the
two distributions to be different at a even larger significance level
($P$(K-S)~$=0.009$).

\begin{figure}
\begin{center}
\includegraphics[width=8cm]{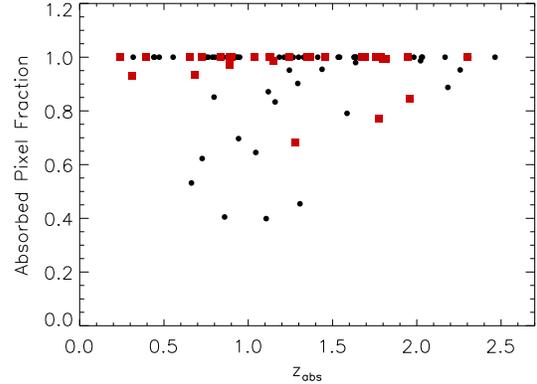}
\caption{Absorbed pixel fraction versus absorption redshift of the systems in our sample. Values of 1 indicate systems where the whole velocity range is absorbed. Symbols are equivalent to those in Figure \ref{ratioMgII}.}
\label{APF}
\end{center}
\end{figure}

Figure \ref{APF} shows the absorbed pixel fraction (APF) distribution
of the {\MgII} \lb2796 absorption profiles, as a function of redshift.
The APF is
defined as the fraction of pixels with detected absorption
over the entire velocity spread of the profile ($\Delta v$),
as used in \citet{Mshar07}. The APF equals one in profiles where there
is absorption over the entire velocity range, and it helps to quantify
the separation between components in the absorption profile whenever
the values are smaller. In Figure \ref{APF} it can be seen that there
is a large concentration of values with continuous coverage (APF equal
1).  The majority (70\%;
61/87) of the absorbers in our sample show a complete absorbed
coverage range, and 80\% (70/87) show a complete or close to complete
(APF~$>0.95$) coverage. We obtain slightly larger percentages if we restrict
our computations to the very strong absorbers alone (72\% and 86\%,
respectively).

As in \citet{Mshar07}, we find differences between the low-redshift
and high-redshift systems.  High-redshift systems tend to show APF
values closer to 1, which indicates that the absorption cover the full
velocity range, while at low redshift it is more likely to find APF
values lower than 1. This is due to a tendency for the
high-redshift systems to have weak components connecting the stronger
absorbing regions.

\begin{figure}
\begin{center}
\includegraphics[width=8cm]{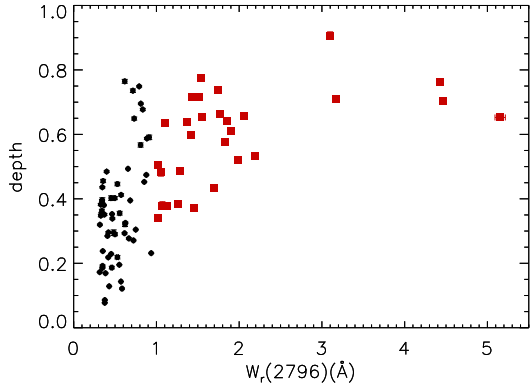}
\caption{Profiles' depth distribution with restframe equivalent width of {\MgII} \lb2796.  Depth is
defined as the fraction of flux absorbed, averaged over the velocity width of a profile.  Symbols are equivalent to those in Figure \ref{ratioMgII}. Moderately strong {\MgII} absorption profiles (black circles) show a larger range of depths than very strong {\MgII} systems (red squares), which show a minimum depth of $\sim$0.3. This is not surprising because there is a correlation between $W_r(\MgII)$ and $\Delta v$.}
\label{depthW}
\end{center}
\end{figure}

\begin{figure}
\begin{center}
\includegraphics[width=8cm]{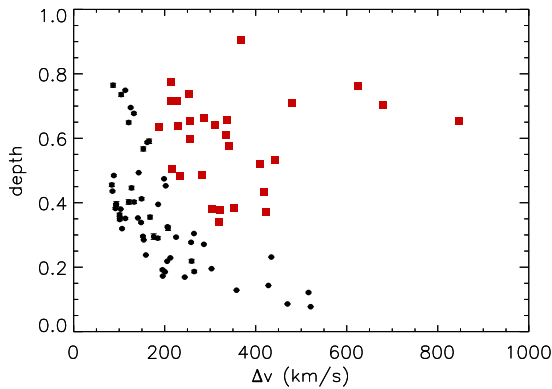}
\caption{Depth versus $\Delta v$ of the profiles in our sample. Symbols are equivalent to those in Figure \ref{ratioMgII}. Moderately strong (black circles) and very strong (red squares) {\MgII} absorbers are confined in different regions of the depth--$\Delta v$ parameter space.}
\label{depthDv}
\end{center}
\end{figure}

\begin{figure}
\begin{center}
\includegraphics[width=8cm]{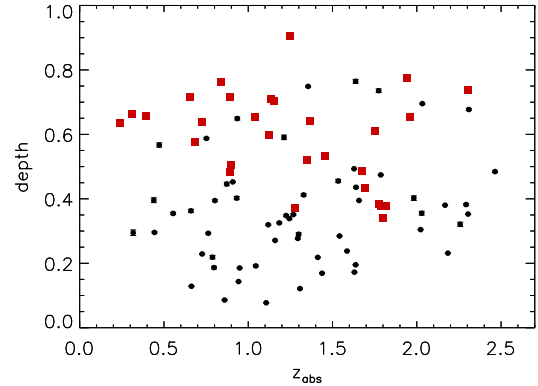}
\caption{Depth versus absorption redshift of the profiles in our sample. Symbols are equivalent to those in Figure \ref{ratioMgII}. We can see that while the moderately strong {\MgII} absorption profiles (black circles) show a similar distribution at low and high redshift, very strong ones (red squares) show evolution with a larger range of depths at high redshifts than at low redshifts.}
\label{depthz}
\end{center}
\end{figure}

Because in the case of very strong absorbers, so many APF values are
close to unity, this measurement does not provide much leverage in
determining kinematic evolution.  To further
quantify the shape of the profiles, and in particular how ``boxy'' or
spread out they are, we define their dimensionless average {\bf depth}
as the fraction of normalized flux that is absorbed, averaged over the
full range of $\Delta v$.  In the case of a completely rectangular
absorber, the depth would be equal to one (since our $W_r$ is defined
from normalized spectra). This averaged depth is equivalent to the
$D$-index as defined in \citet{Ellison06b}, and posteriorly reviewed
in \citet{Ellison09}, except that we include the whole velocity range
($\Delta v$) in the average. 
We present this depth measurement versus some other properties of
our sample in Figures \ref{depthW}, \ref{depthDv}, and \ref{depthz}.
Figure \ref{depthW} and Figure \ref{depthDv} show profile depths
versus $W_r$(\MgII) and $\Delta v$, respectively.
There is an envelope at the lower end of the depth distribution
which depends on the $W_r(\MgII)$ value. This is regulated by
the typical maximum $\Delta v$ value for the profiles. In other words,
we do not find shallow profiles with large $W_r(\MgII)$ values
(see Figures \ref{mosaic} and \ref{Fig13}), which would result
from very large $\Delta v$ values and would lead to a departure from the
$W_r(\MgII)$--$\Delta v$ correlation. Only the
several profiles with the largest $W_r(\MgII)$ have
$\Delta v \gsim 500$--$600$~{\kms}. Thus, a given
equivalent width can only result from a depth larger
than some minimum value.
Completely "boxy" profiles are also rare, since weaker subsystems are
common; because of this and the gradual recovery of profiles to
meet the continuum, there is also a natural high limit for the depth value.
Figure \ref{depthDv} shows that moderately strong and very strong {\MgII}
absorbers occupy a different region of the depth--$\Delta v$ parameter
space. This is also partly due to the correspondence between larger $W_r$
and larger $\Delta v$.  Once saturation is present the only way
to increase the $W_r$ is through increasing the $\Delta v$ of the
profile.

Figure \ref{depthz} shows the distribution of depth with absorption
redshift. There is a clear evolution for the very strong {\MgII}
absorption profiles, which is not present for the moderately strong
ones.  An Anderson-Darling test results in a probability of only $P=$~0.03
that the depths of the low redshift subsample of very strong absorbers
are drawn from the same population as those of the high redshift
subsample. The distribution of depths of the moderately strong (black
circles) and very strong absorbers (red squares) at low redshift is
significantly more different ($P$(K--S)=~4.e-7) than the distribution
of the two subsamples at high redshift ($P$(K--S)=~0.02). Because the
depth is influenced both by $W_r$(\MgII) and $\Delta v$, evolution of
$W_r$(\MgII) would play a role into depth evolution. However, in
Figure \ref{zadist} we already showed that there is no significant
evolution of $W_r$(\MgII) with redshift in our sample, thus changes in
$\Delta v$ must dominate. Another way to consider the trend of depth
evolution in Figure \ref{depthz} is to note that the cluster of
very strong absorbers with depth values $\sim 0.4$ in Figure \ref{depthDv} are all
in our high redshift sample.  The presence of small depths at high
redshift, but not at low redshift, is reminiscent of the evolution of
$W_r(\FeII)/W_r(\MgII)$ shown in Figures \ref{ratioz}, and we will
consider this connection in \S~\ref{sec:4}.

\section {Summary and Discussion}
\label{sec:4}

In a large sample (81) of high resolution VLT/UVES spectra, we
have found 58 moderately strong (0.3~$<W_r<$~1 \AA) and 29 very strong
($W_r>$~1 \AA) {\MgII} absorption systems, and searched for accompanying
{\FeII} and {\MgI} absorption. We have investigated the profile shape evolution of {\MgII}
as well as the evolution of the $W_r(\FeII)/W_r(\MgII)$ ratio. Here we summarize the
main results of this study.

\begin{itemize}
\item For moderately strong {\MgII} absorbers, the ratio $W_r(\FeII)/W_r(\MgII)$, does
not evolve significantly from $z\sim 2.5$ to $z\sim 0.2$.
\item On the contrary, very strong {\MgII} absorbers show evolution in $W_r(\FeII)/W_r(\MgII)$, with a
deficit of small values at low redshift.
\item Moderately strong {\MgII} absorber profiles do not evolve significantly from
$z\sim 2.5$ to $z\sim 0.2$, as quantified by profile velocity spread and depth.
\item However, very strong {\MgII} absorbers at low redshift all have relatively
high profile depth values, while those at high redshift have a wider range of
values (more similar to the range for moderately strong systems).
\end{itemize}

\subsection{Possible Causes for the Evolution of Very Strong {\MgII} Absorbers}
\label{sec4.1}

It is not surprising to find an evolution in the population of very
strong {\MgII} systems. \citet{Nestor05} already showed an excess of
these systems at high redshift, relative to absorbers with smaller
equivalent widths. Among the very strong {\MgII} absorber population,
the deviations from cosmological evolution are larger as $W_r(\MgII)$
increases (\citealt{Nestor05}).

Given the observed evolution in the $W_r(\FeII)/W_r(\MgII)$ ratio for
the class of very strong {\MgII} absorbers, shown in Figure
\ref{Fig13}, we consider several plausible reasons.  These
include: 1) a changing ionization rate due to evolution of the
extragalactic background; 2) a change in the level of
$\alpha$--enhancement due to changing contributions of Type II and
Type Ia supernovae to the absorbing gas; 3) a change in the kinematics
of the profiles, including different levels of saturation and velocity
spreads which affect the equivalent widths of {\FeII} and {\MgII}
differently. Here we will consider the expected effect of each of
these factors.

The parameter space of very strong {\MgII} absorbers shows large
$W_r(\FeII)/W_r(\MgII)$ at both low and high redshift, but small
$W_r(\FeII)/W_r(\MgII)$ only at high redshift (see Figs.~\ref{ratioz}
and \ref{Fig13}).  The profiles of very strong {\MgII} absorbers
(those with $W_r>$1~{\AA}) are almost always saturated in {\MgII}, and
sometimes saturated even in the weakest accessible {\FeII} transition
as well (see Fig.~\ref{mosaic}).  We should first consider this when
interpreting the observed absence of low $W_r(\FeII)/W_r(\MgII)$
values in this population at low redshift.  As shown in \ref{3.1},
Figure \ref{Fig13} indicates that there is an evolution in the {\MgII}
profile kinematics that is relevant for our interpretation of the
$W_r$(\FeII)/$W_r(\MgII)$ evolution for very strong {\MgII} absorbers.
At low redshift, many of the systems show ``boxy'' profiles and they
tend to have fewer, if any, satellite clouds around their compact,
``boxy'' portions.  Therefore, they do not have a significant range of
velocity that is unaffected by saturation in {\MgII}.  A typical
example of these ``boxy'' profiles is the $z =$~0.2378 system toward
Q0952+0179, shown in Figure \ref{profiles} (also see Fig.~\ref{mosaic}
and Fig.~\ref{Fig13}).  Because the weaker {\FeII} \lb2374 and {\FeII}
\lb2587 transitions are not strongly saturated for many components,
the {\FeII} equivalent widths are less affected by saturation effects.

Also, it is important to note that we are only comparing the equivalent
widths of the {\MgII} and {\FeII} transitions, and not their column
densities.  In the case of a saturated profile, of course it is not
possible to measure the column density, and we should interpret the
equivalent width as a lower limit on how much material is present.
For the $W_r(\FeII)/W_r(\MgII)$ ratio, in cases where the {\MgII} is
saturated and the {\FeII} is not, our values are therefore upper
limits.  This alone could lead to the absence of very strong {\MgII}
absorbers with small $W_r(\FeII)/W_r(\MgII)$ values.  The question
then is why very strong {\MgII} absorbers at high redshift
do often have small values of $W_r(\FeII)/W_r(\MgII)$, while those
at low $z$ do not.

The answer can be seen by examining the {\MgII} profile shapes of the
high redshift systems in the lower right quadrants of Figures
\ref{ratioz} and \ref{Fig13}.  These profiles show components more
spread out in velocity space, with several subsystems within the
absorption profile.  Because of this kinematic spread, even if the
column densities and numbers of components are the same as for the low
redshift systems, we will see less saturation.  This is evident in the
existence of some low values of profile depth (see Figure
\ref{depthz}) among the high redshift very strong {\MgII}
absorbers. Several ``boxy'' saturated {\MgII} profiles are apparent
among the high $z$ systems as well, and these have large values of
$W_r(\FeII)/W_r(\MgII)$, as did all of the systems at lower redshift.
If the {\MgII} profile suffers from significant saturation, the
$W_r(\FeII)/W_r(\MgII)$ value is an overestimate and it is impossible
for it to be small.  In order that a very strong {\MgII} absorber
not have significant saturation it must be kinematically spread, and
such kinematics are apparently more common at high redshift than at
low redshift. 

This is consistent with the conclusions of \citet{Mshar07} on the
evolving kinematics of a smaller sample of very strong {\MgII}
absorbers.  That study found that complex systems with many components
spread in velocity are more common at high redshift than at low
redshift.  For example, profiles typically associated with superwinds
(\citealt{Bond01}; \citealt{Ellison03}) show large velocity spreads.
The excess of very strong {\MgII} absorbers at high redshift, found in
\citet{Nestor05}, may be related to our findings if the excess systems
correspond to outflows/superwinds, and are thus evolving together with the star
formation rate in galaxies.

Returning to the three effects that could affect evolution of the
$W_r(\FeII)/W_r(\MgII)$ ratio in very strong {\MgII} systems, we have
found that kinematic evolution of this population could explain the
observed trend.  This does not necessarily mean that ionization and
abundance pattern changes are not taking place as well, just that we
are not able to get an indication of the ionization parameter or
abundance pattern for the saturated systems.  In other words, some of
the systems with ``boxy'' profiles in {\MgII} both at high and low
redshift, could have higher/lower ionization parameters or could be
$\alpha$-enhanced/depleted relative to the others, but would still not
have small $W_r(\FeII)/W_r(\MgII)$ ratios relative to others because
of saturation of the {\MgII}. At higher redshift, however, it seems
likely that systems with large kinematic spread and small values of
$W_r(\FeII)/W_r(\MgII)$ (bottom right quadrant of Figure \ref{Fig13})
are different than the ``boxy'' profiles systems (top right quadrant
of Figure \ref{Fig13}) that also appeared in this redshift range.  The
strong {\MgII} absorbers, with 0.3~{\AA}~$< W_r(\MgII) <$~1~{\AA},
also have the full range of $W_r(\FeII)/W_r(\MgII)$, but not all of
them with large values are strongly saturated in {\MgII}.

It is natural to expect an absorber population to be more ionized at
higher redshift, because of the evolution of the extragalactic
background radiation.  This is due both to the changes in the ionizing
extragalactic background radiation and due to the more intense local
radiation field in starbursts, both of which are a consequence of the
higher global star formation rate at higher redshift.  Even without a
contribution from local star formation, the photon number density,
$\log n_{\gamma}$ changes from $-4.7$~{\cc} at $z=2$ to $-5.7$~{\cc}
at $z=0.3$, which will lead to a significant difference in the
ionization parameter affecting our {\MgII} absorbing clouds.  In
particular, $N(\FeII)/N(\MgII)$ is relatively constant for ionization
parameter $\log U < -4.5$ but decreases rapidly from $\log U = -4$ up
to higher values. At $z=2$, a cloud with $n_H = 0.1$~{\cc} would have
$\log U = -3.7$, while it would have $\log U = -4.7$ at $z=0$. If
instead the {\MgII} cloud has lower densities, for example $n_H =
0.01$~{\cc}, it would result in $\log U = -2.7$ at $z=2$, and $\log U
= -3.7$ at $z=0$.  While the difference of $N(\FeII)/N(\MgII)$ between
low and high redshift systems would be barely noticeable in the case
of $n_H = 0.1$~{\cc}, clouds of densities of $n_H = 0.01$~{\cc} would
have significantly lower $N(\FeII)/N(\MgII)$ at high redshift than at
low redshift.  Thus, the systems with small $N(\FeII)/N(\MgII)$
values would result from systematically lower density clouds. A larger range
of densities (and thus $\log U$) would lead to ionization of the
{\FeII} and would give rise to small $N(\FeII)/N(\MgII)$ at high
redshift, and thus we would expect more systems with small
$N(\FeII)/N(\MgII)$ at high redshift.  Although this clearly matches
the observed behavior for very strong {\MgII} absorbers, we cannot
measure this effect since it is disguised by the saturated {\MgII}
profiles in many of the systems.

Similarly, $\alpha$--enhanced systems are expected to be more common
at higher redshift, where the star formation rate is higher.  It takes
time for a stellar population to produce Type Ia supernovae, leading
to about a 1 billion year delay before the iron abundance would be
elevated to the solar value, relative to magnesium.  As will be
discussed further in \ref{sec4.2}, \citet{Narayanan07} noted that this
delay could be a significant contributing factor to the absence of
weak {\MgII} absorbers with large $W_r$(\FeII)/$W_r(\MgII)$.  The
larger kinematic spreads of some of the high-redshift very strong
{\MgII} absorbers may be related to enhanced star formation/starburst
and wind activity at high redshift.  Thus, these systems would be
expected to be the $\alpha$-enhanced systems with small
$W_r$(\FeII)/$W_r(\MgII)$ values.  It is plausible that there would be
fewer such $\alpha$--enhanced systems at low redshift.  Also, a
stronger depletion of {\Fe} relative to {\Mg} would result in a
smaller $W_r(\FeII)/W_r(\MgII)$. It is known that {\Mg} returns to the
gas phase faster than {\Fe} (\citealt{Fitzpatrick96}). Indeed, mild
and high {\Fe} depletion has been previously observed in some DLA and
sub-DLA systems, both at low and high redshift (e.g.,
\citealt{Ledoux02}; \citealt{Meiring08}; \citealt{Noterdaeme10};
\citealt{Meiring11}). Like ionization effects, both
$\alpha$--enhancement and depletion of {\Fe} contribute in the right
direction to lead to a relative absence of very strong {\Mg}
absorption in the lower right quadrant in the
$W_r$(\FeII)/$W_r(\MgII)$ vs. $z$ diagram.  However, we cannot tell if
the lower left quadrant is vacant because of the lack of
$\alpha$--enhancement/greater depletion of {\Fe} at low redshift, or just
because low redshift systems preferentially have kinematics that lead
to saturation of {\MgII} and {\FeII}.

We conclude that, in the case of very strong {\MgII} absorbers, low
values of $W_r$(\FeII)/$W_r(\MgII)$ at high redshift could be due to a
larger ionizing radiation field or to $\alpha$--enhancement of a
larger fraction of systems, but the effects cannot be quantified in
these mostly saturated systems.  We also cannot be certain of the
cause of the absence of low $W_r(\FeII)/W_r(\MgII)$ values at low
redshift, because all the low redshift systems in our
sample have significant saturation of {\MgII} and {\FeII}. The significant
evolution that we have detected, therefore, is the evolution of the
kinematics of the {\MgII} profiles, from a mixture of saturated, flat
bottom profiles and kinematically spread components at high redshift,
to only predominantly saturated, flat bottom profiles at low redshift.

This is quite consistent with several recent studies of the nature of DLAs
and sub-DLAs, which are coincident with the population of very strong
{\MgII} absorbers.  Since very strong {\MgII} absorbers are selected by
an equivalent width criterion, they are not necessarily all among the
highest column density absorbers, and thus are not all DLAs.  The equivalent
width of {\MgII} can be very large either because the total column density
of material is very large (thus it would likely be a DLA), or because
the kinematic spread of components is large enough that the equivalent
widths add up to a large total (in which case the system might be a DLA, but may also be
a sub-DLA or even a Lyman limit system).  

Indeed, the populations of DLAs and sub-DLAs have important intrinsic
differences.  \citet{Kulkarni10} show that DLAs have systematically
lower metallicities than sub-DLAs (based on {\Zn} measurements), and
furthermore that they have a smaller velocity spread than sub-DLAs.
\citet{Kulkarni10} suggests that the differences in metallicities
between DLAs and sub-DLAs could be due to different star formation
histories, where the galaxies observed as sub-DLAs undergo more
rapid star formation.  Both \citet{Kulkarni10} and \citet{Nestor11}
suggest that DLAs are often dense regions in ordinary galaxies, while
sub-DLAs are of higher metallicity and extended kinematics such as one
would expect from a superwind.  All these results would be consistent
with the boxy profiles that we see in {\MgII} tending to be DLAs,
while the kinematically spread systems that are more common at higher
redshift, would tend to be sub-DLAs, where the star formation is more
likely to happen.  Then, our finding that small
$W_r$(\FeII)/$W_r(\MgII)$ values for very strong {\MgII} absorbers are
only present at high redshifts could be due to both $\alpha$-enhancement
and kinematics, expected from outflows/superwinds.

Another related factor that would condition the shape of the {\MgII}
profiles may be the association of the subset of very strong absorbers
with galaxy over-density regions such as groups or clusters where
interactions and mergers are common. \citet{Nestor07} found that field
imaging of a subset of the strongest {\MgII} absorbers
($W_r(\MgII)>2.7$~{\AA}) at low redshift ($0.42 < z < 0.84$) indicates
that pairs of galaxies, often with distorted morphologies, and
starburst related phenomena, are coincident with these {\MgII}
systems. Although, this redshift range is lower compared to the
high--$z$ very strong systems in our sample, interactions would be
only more frequent at high redshift.  If these very strong {\MgII}
absorption systems are selecting group environments, star formation is
also going to be higher compared to the average for that redshift
(e.g.  \citealt{Kennicutt87}; and more recently \citealt{Wong11}),
which would push this subset of absorbing gas in the direction towards
lower $W_r(\FeII)/W_r(\MgII)$ ratios.  We find that our results are
consistent with this scenario.  In our sample, for one of the
strongest {\MgII} systems (Q0002-422, $z =$~0.837), \citet{Yanny90}
have identified several emission line sources within 100 kpc, which
indicates that there is probably a group environment (also see
\citealt{York91}).  Similarly, the very strong {\MgII} system in the
line of sight of Q0453-423 at $z =$~0.726 also has multiple galaxies
associated with it (\citealt{Yanny90}).

\subsection{Comparison to the Evolution of Weak {\MgII} Absorbers}
\label{sec4.2}

\begin{figure}
\begin{center}
\includegraphics[height=6.3cm]{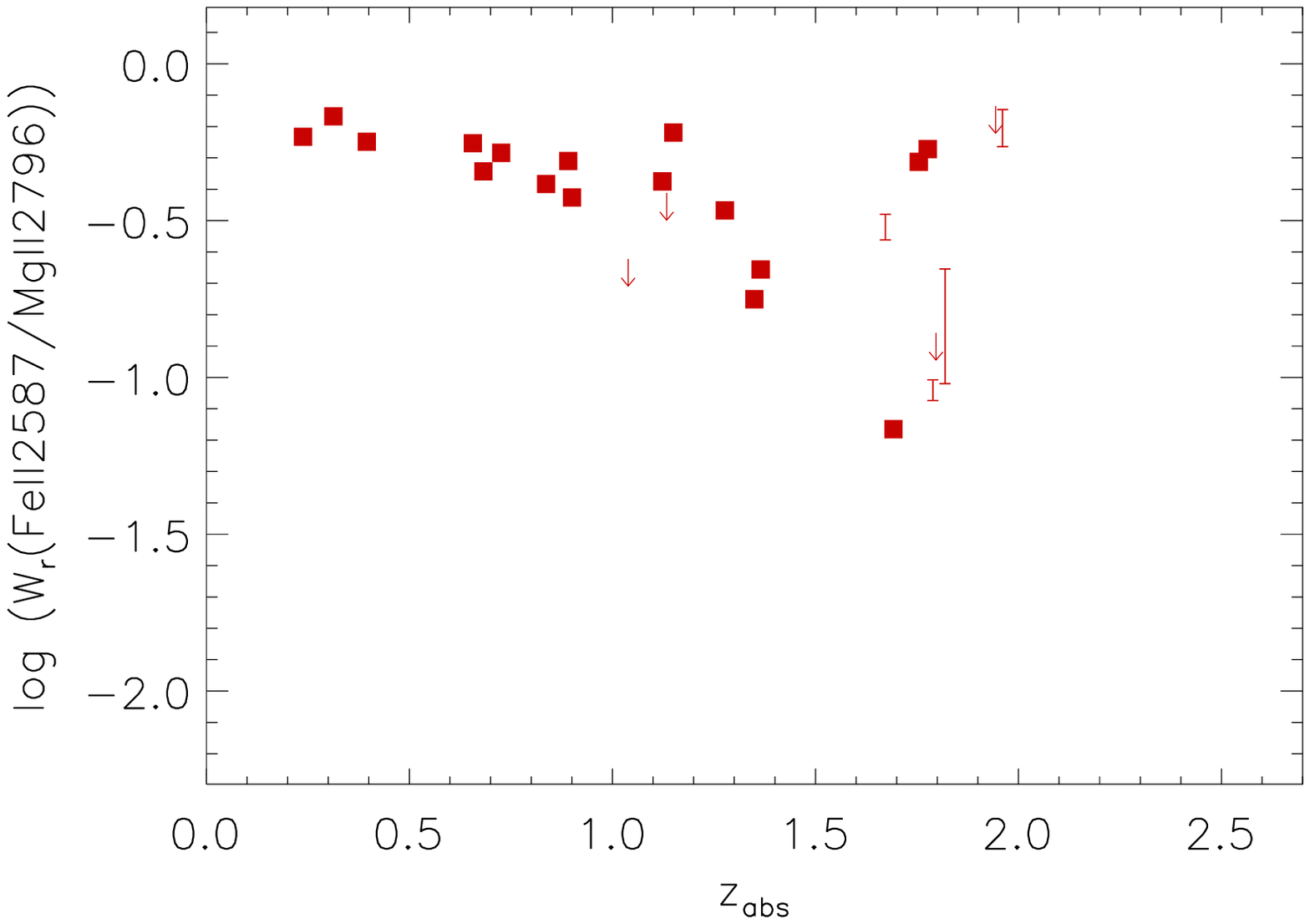}
\includegraphics[height=6.3cm]{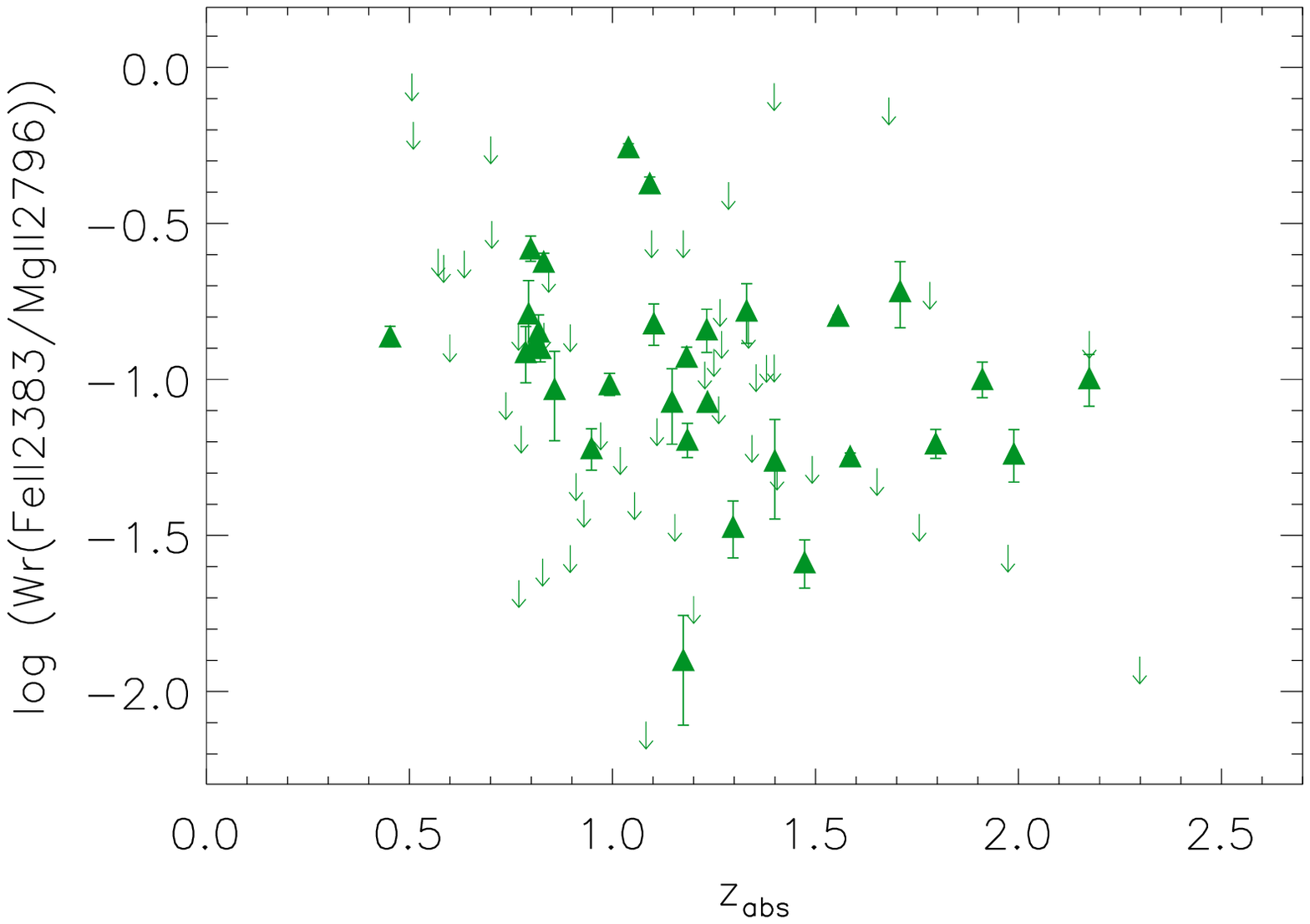}
\caption{Ratio of $W_r(\FeII)/W_r(\MgII)$ versus absorption redshift ($z_{abs}$) for the very strong {\MgII} systems (top) and weak {\MgII} systems (bottom). Symbols are equivalent to those in Figure \ref{ratioMgII}. Two opposite trends are present. There is an absence of very strong absorbers with low ratios at low redshift (top) and an absence of weak absorbers with large ratios at high redshift (bottom). In the case of the very strong absorbers, we chose {\FeII} \lb2587, because it was less often saturated than the {\FeII} \lb2383 and \lb2600 transitions, and was covered more often than the weaker {\FeII} \lb2374 transition. For weak {\MgII} absorbers, {\FeII} \lb2383 was chosen because, as the strongest {\FeII} transition, it is mostly likely to be detected.}
\label{ratioz2}
\end{center}
\end{figure}

The evolution of weak systems and their $W_r(\FeII)/W_r(\MgII)$ ratio
was presented in \citet{Narayanan07} and \citet{Narayanan08}. Figure
\ref{ratioz2} shows the evolution of the $W_r$ ratio for very strong {\MgII}
absorption systems (red squares -- top) and the weak
{\MgII} systems (green triangles -- bottom) from
\citet{Narayanan07}. For very strong absorption (top) $W_r(\FeII)$
is measured in the {\FeII} \lb2587 transition because it is weak
enough not to be saturated,
but is the most often observed {\FeII} transition in our
sample.  Weak {\MgII} absorption is accompanied by even weaker
{\FeII} troughs, and since {\FeII} \lb2383 is the strongest
transition, it is the mostly likely to be detected and the one
displayed in Figure \ref{ratioz2}.\footnote{Table 1 in
\citealt{Narayanan08} lists the weak absorbers $W_r(\MgII)$ and
$W_r(\FeII2383)$ measurements and limits.}  Weak absorbers
reflect a different trend than very strong absorbers: we find that
large $W_r(\FeII)/W_r(\MgII)$ values are uncommon at higher
redshifts, i.e., weak {\MgII} systems show a deficit as well, but
in the opposite quadrant of the $W_r(\FeII)/W_r(\MgII)$ -- $z$
parameter space.

\citet{Narayanan07} considered the possible causes of the evolution of
weak {\MgII} absorbers.  Since saturation of these weak profiles is
fairly rare, the cause of the evolution could be 1) changing
ionization or 2) abundance pattern changes in the population of gas clouds
that produce these absorbers. In the first case, because of the known
evolution of the extragalactic background radiation (higher at higher
redshifts) and since {\FeII} is more readily ionized than {\MgII}, we
would need higher density absorbers to populate the large
$W_r(\FeII)/W_r(\MgII)$ high redshift quadrant, than we would at low
redshift.  In the second case, a delay between the Type II and Type Ia
SNe, i.e., an $\alpha$--enhancement, might help explain the absence of
large $W_r(\FeII)/W_r(\MgII)$ values at high redshifts.  There may not
yet have been time for Type Ia supernovae to be active in weak {\MgII}
absorbing clouds at high redshift, which themselves tend to be less
common at $z>1.4$ than at lower redshifts (see \citealt{Narayanan08};
\citealt{Narayanan07}; \citealt{Lynch06}).

In fact, the evolution of the very strong {\MgII} absorbers and the
weak {\MgII} absorbers, though in the opposite sense, may be
related. Weak {\MgII} absorbers with $N(\FeII) \sim N(\MgII)$ require
an abundance pattern influenced by Type Ia supernovae, and their
relative absence at $z>1.4$ implies that they are related to a
population of gas clouds that is not common at high redshifts.  As
described in \citet{Narayanan08}, this could either be because they
are related to dwarf galaxies, in which star formation peaks later, or
to a rarity of weak {\MgII} clouds because halos are more crowded so
they are more commonly combined to form strong {\MgII} absorbers.  At
least some weak {\MgII} absorbers are produced by the same gas clouds
that produce the ``satellite clouds'' of strong {\MgII} absorbers,
simply viewed from a different vantage point.  These are also likely
to be analogous to Milky Way high velocity clouds, at various
redshifts (\citealt{Richter09}).

What we have seen here is that some fraction of very strong {\MgII}
absorbers, those that are related to sub-DLAs and to star formation
activity, have components spread in velocity space.  Their
$W(\FeII)/W(\MgII)$ values are consistent with $\alpha$--enhancement,
just as are all of the weak {\MgII} absorbers in the same redshift
range.  Since satellites around very strong {\MgII} absorbers are
analogs to isolated weak {\MgII} absorbers, and since these are common
in high redshift sub-DLAs, the picture of $\alpha$--enhancement in
both populations is consistent.

\subsection{Conclusions and Future Work}
\label{sec4.3}

The main result of this study is the evolution of the population of
very strong {\MgII} absorbers ($W_r({\MgII}) > 1.0$~{\AA}) over the
redshift range, $0.2 < z < 2.5$.  There is a significant absence of
small $W_r({\FeII})/W_r({\MgII})$ values at low redshift, as compared
to the full range apparent at high redshift.  A major contributing factor
to this change is cleary the kinematic evolution of the profiles of the very
strong absorbers.  At low redshift, {\MgII} presents simple, boxy, saturated
profiles, while at high redshift there are
also {\MgII} profiles with multiple components, some unsaturated, that
are spread over a larger range of velocity.  Thus at high redshift, it
is more likely that some of the {\FeII} profiles will also be unsaturated, so
that small $W_r({\FeII})/W_r({\MgII})$ values can arise.

Other factors could also contribute to the absence of 
small $W_r({\FeII})/W_r({\MgII})$ values at low redshift.
Most notably a small ratio could be indicative of an
$\alpha$--enhanced abundance pattern, which would be characteristic
of a young stellar population, such as one contributing to active
galactic outflows.  The absence of small ratios of {\FeII} to
{\MgII} at low redshift, is thus consistent with the idea
that outflows are more active at high redshift.

The key to testing the hypothesis that abundance pattern is evolving,
in addition to the known evolution in the kinematics, is to determine
if the very strong {\MgII} absorbers at $z>1.2$ that have small
$W(\FeII)/W(\MgII)$ are truly $\alpha$--enhanced.  This requires a
separation of saturation effects from the effect of changing abundance
patterns.  Transitions with lower oscillator strengths should be
studied to facilitate this separation.  There are numerous {\FeII}
transitions with various oscillator strengths in the rest frame
$\lambda=1050$--$2600$~{\AA} regime.  Since other {\MgII} transitions
are too weak to be detected, even in these very strong systems, we
will need to consider other $\alpha$--elements for comparison,
e.g. {\SiII}, which has several promising transitions in the rest
frame far--UV, an ionization potential similar to {\MgII}, and similar
dust abundances at least in the Galactic disk and the ISM
(\citealt{Cartledge06}; \citealt{Whittet10}).  High resolution quasar
spectra with wide wavelength coverage (optical and UV) are thus
essential to assessing evolving chemical abundance processes over
cosmic time.

\section*{Acknowledgments}

This research was funded by NASA under grant NAG5-6399 NNG04GE73G, and
by the National Science Foundation (NSF) under grant AST-04-07138 and
through the REU program. P.R.H. would like to thank those who
approached us during talks and posters, because those discussions
helped shape the explanations in this paper.

\bibliographystyle{mn2e}
\bibliography{bibliography}

\clearpage
\newpage

\onecolumn

\begin{center}
\begin{longtable}{llllllll}

\caption{Rest-frame Equivalent Widths for Target Transitions} \label{data} \\

\hline \hline \\[-2ex]
   \multicolumn{1}{c}{{QSO}} &
   \multicolumn{1}{c}{{$z_{abs}$}} &
   \multicolumn{1}{c}{{$W_{\rm r}(\MgII)$ (\AA)}} &
   \multicolumn{1}{c}{{$W_{\rm r}(2374)$ (\AA)}} &
   \multicolumn{1}{c}{{$W_{\rm r}(2383)$ (\AA)}} &
   \multicolumn{1}{c}{{$W_{\rm r}(2587)$ (\AA)}} &
   \multicolumn{1}{c}{{$W_{\rm r}(2600)$ (\AA)}} &
   \multicolumn{1}{c}{{$W_{\rm r}(2853)$ (\AA)}} \\[0.5ex] \hline
   \\[-1.8ex]
\endfirsthead

\multicolumn{3}{c}{{\tablename} \thetable{} -- Continued} \\[0.5ex]
  \hline \hline \\[-2ex]
  \multicolumn{1}{c}{{QSO}} &
   \multicolumn{1}{c}{{$z_{abs}$}} &
   \multicolumn{1}{c}{{$W_{\rm r}(\MgII)$ (\AA)}} &
   \multicolumn{1}{c}{{$W_{\rm r}(2374)$ (\AA)}} &
   \multicolumn{1}{c}{{$W_{\rm r}(2383)$ (\AA)}} &
   \multicolumn{1}{c}{{$W_{\rm r}(2587)$ (\AA)}} &
   \multicolumn{1}{c}{{$W_{\rm r}(2600)$ (\AA)}} &
   \multicolumn{1}{c}{{$W_{\rm r}(2853)$ (\AA)}} \\[0.5ex] \hline

  \\[-1.8ex]
\endhead

  \multicolumn{3}{l}{{Continued on Next Page\ldots}} \\
\endfoot

 \\[-1.8ex] \hline \hline
\endlastfoot

Q0001-2340 &  0.949 &  0.347$\pm$ 0.001 & $<$0.0015 & $<$ 0.047 & $<$0.033 & 0.0397$\pm$0.0009 & 0.0090$\pm$0.0007\\
  &  1.586 &  0.352$\pm$ 0.002 & $<$0.011 & 0.0484$\pm$0.0005 & 0.0144$\pm$0.0004 & 0.0420$\pm$0.0007 &  0.017$\pm$ 0.001\\
  &  2.184 &  0.937$\pm$ 0.002 &  0.021 -  0.035 &  0.193$\pm$ 0.002 & $<$ 0.22 & $<$ 0.38 & $<$ 0.86\\
Q0002-0422 &  0.837 &  4.431$\pm$ 0.002 &  0.874 -  0.966 &  3.055$\pm$ 0.002 &  1.833$\pm$ 0.003 &  2.995$\pm$ 0.002 &  1.586$\pm$ 0.002\\
  &  1.542 &  0.408$\pm$ 0.001 & 0.0143$\pm$0.0009 & 0.0718$\pm$0.0006 & 0.0238$\pm$0.0009 & 0.0664$\pm$0.0006 & $<$0.038\\
  &  2.168 &  0.366$\pm$ 0.001 & 0.0106$\pm$0.0006 &  0.084$\pm$ 0.001 & $<$ 0.076 & $<$ 0.12 & ...\\
  &  2.302 &  1.738$\pm$ 0.002 &  0.203$\pm$ 0.001 &  0.668$\pm$ 0.001 & ... & ... & ...\\
  &  2.464 &  0.398$\pm$ 0.002 & $<$0.014 & $<$0.036 & ... & $<$ 0.061 &  0.008$\pm$ 0.002\\
Q0010-0012 &  1.212 &   0.91$\pm$  0.01 &  0.102$\pm$ 0.003 &  0.344$\pm$ 0.008 &  0.160$\pm$ 0.003 &   0.38$\pm$  0.01 &  0.129$\pm$ 0.009\\
Q0013-0029 &  2.029 &  0.556$\pm$ 0.009 &  0.556$\pm$ 0.009 &  0.122$\pm$ 0.004 &  0.035$\pm$ 0.005 &  0.115$\pm$ 0.005 &  0.064$\pm$ 0.007\\
Q0055-0269 &  1.268 &  0.372$\pm$ 0.003 & $<$0.0039 & $<$ 0.19 &  0.017$\pm$ 0.002 &  0.050$\pm$ 0.003 &  0.037$\pm$ 0.003\\
  &  1.534 &  0.356$\pm$ 0.004 &  0.106$\pm$ 0.003 &  0.211$\pm$ 0.002 &  0.151$\pm$ 0.002 &  0.230$\pm$ 0.002 &  0.021$\pm$ 0.002\\
Q0100+1300 &  1.797 &  1.014$\pm$ 0.007 & ... & ... & $<$ 0.14 &  0.196$\pm$ 0.004 &  0.101$\pm$ 0.004\\
  &  2.309 &  0.835$\pm$ 0.004 &  0.044$\pm$ 0.001 &  0.013$\pm$ 0.002 & ... & ... & ...\\
Q0109-3518 &  1.350 &  1.980$\pm$ 0.002 & ... &  0.839$\pm$ 0.002 &  0.352$\pm$ 0.002 &  0.790$\pm$ 0.002 &  0.293$\pm$ 0.002\\
Q0112+0300 &  1.245 &   3.10$\pm$  0.02 &   1.02$\pm$  0.02 &   2.19$\pm$  0.02 & ... & ... &  1.070$\pm$  0.02\\
Q0122-0380 &  0.444 &  0.419$\pm$ 0.004 & $<$0.0041 &  0.041$\pm$ 0.003 & $<$ 0.017 & $<$ 0.16 &  0.045$\pm$ 0.004\\
  &  0.860 &  0.376$\pm$ 0.004 & 0.0080$\pm$0.0009 &  0.067$\pm$ 0.002 & 0.0204$\pm$0.0009 &  0.059$\pm$ 0.002 &  0.030$\pm$ 0.002\\
  &  1.244 &  0.468$\pm$ 0.003 & 0.1107$\pm$0.0009 &  0.227$\pm$ 0.002 & ... & ... &  0.109$\pm$ 0.001\\
Q0130-4021 &  0.931 &  0.497$\pm$ 0.006 & ... & ... & $<$0.0078 &  0.057$\pm$ 0.006 & $<$ 0.40\\
  &  0.935 &  0.731$\pm$ 0.005 & ... & ... &  0.152$\pm$ 0.005 &  0.289$\pm$ 0.004 & $<$ 0.30\\
Q0136-0231 &  0.802 &  0.342$\pm$ 0.003 &  0.035$\pm$ 0.004 &  0.178$\pm$ 0.004 &  0.081$\pm$ 0.002 &  0.185$\pm$ 0.003 &  0.058$\pm$ 0.004\\
  &  1.184 &  0.625$\pm$ 0.005 &  0.056$\pm$ 0.005 &  0.207$\pm$ 0.003 &  0.081$\pm$ 0.005 &  0.162$\pm$ 0.003 &  0.038$\pm$ 0.003\\
  &  1.294 &  0.666$\pm$ 0.006 &  0.071$\pm$ 0.003 & $<$ 0.32 &  0.147$\pm$ 0.003 &  0.233$\pm$ 0.004 &  0.164$\pm$ 0.006\\
Q0151-4326 &  0.663 &  0.429$\pm$ 0.002 &  0.028$\pm$ 0.001 &  0.131$\pm$ 0.001 & $<$ 0.09 &  0.125$\pm$ 0.001 &  0.020$\pm$ 0.001\\
Q0237-0023 &  1.365 &  1.856$\pm$ 0.001 &  0.206$\pm$ 0.001 &  0.885$\pm$ 0.001 &  0.410$\pm$ 0.001 &  0.870$\pm$ 0.001 &  0.268$\pm$ 0.001\\
  &  1.637 &  0.551$\pm$ 0.002 & $<$0.018 & $<$ 0.19 & 0.0400$\pm$0.0009 &  0.125$\pm$ 0.001 &  0.049$\pm$ 0.001\\
  &  1.657 &  0.683$\pm$ 0.001 & 0.0101$\pm$0.0006 &  0.107$\pm$ 0.001 & ... & $<$ 0.20 &  0.029$\pm$ 0.001\\
  &  1.672 &  1.283$\pm$ 0.001 &  0.256$\pm$ 0.001 &  0.545$\pm$ 0.001 &  0.354 -  0.426 &  0.551$\pm$ 0.001 & $<$ 0.76\\
CTQ0298 &  1.039 &  1.554$\pm$ 0.006 & $<$ 0.90 & $<$ 0.82 & $<$ 0.37 &  0.736$\pm$ 0.003 &  0.176$\pm$ 0.008\\
Q0300+0048 &  0.892 &   1.05$\pm$  0.01 &   0.18 &   0.30 &   0.07 &   0.50$\pm$  0.01 &  0.330$\pm$  0.01\\
Q0328-0272 &  0.788 &   0.53$\pm$  0.01 &  0.016$\pm$ 0.003 &   0.21$\pm$  0.01 &  0.057$\pm$ 0.009 &  0.109$\pm$ 0.007 & ...\\
  &  1.123 &   1.42$\pm$  0.01 &  0.402$\pm$ 0.008 &  0.912$\pm$ 0.008 &  0.599$\pm$ 0.007 &  0.946$\pm$ 0.008 & ...\\
  &  1.299 &  0.500$\pm$ 0.008 &  0.045$\pm$ 0.006 &  0.098$\pm$ 0.007 &  0.028$\pm$ 0.005 &  0.048$\pm$ 0.005 &  0.025$\pm$ 0.004\\
  &  1.307 &  0.584$\pm$ 0.009 & $<$ 0.043 &  0.078$\pm$ 0.006 &  0.023$\pm$ 0.005 &  0.045$\pm$ 0.006 & ...\\
Q0329-0385 &  0.763 &  0.615$\pm$ 0.005 &  0.021$\pm$ 0.002 &  0.156$\pm$ 0.003 &  0.103$\pm$ 0.005 &  0.166$\pm$ 0.005 &  0.055$\pm$ 0.003\\
  &  1.438 &  0.385$\pm$ 0.004 & ... & ... &  0.020 -  0.040 &  0.056$\pm$ 0.003 &  0.039$\pm$ 0.002\\
Q0429-4901 &  0.554 &  0.335$\pm$ 0.003 &  0.020$\pm$ 0.003 &  0.080$\pm$ 0.004 &  0.015$\pm$ 0.002 &  0.085$\pm$ 0.004 & $<$0.0039\\
  &  1.119 &  0.317$\pm$ 0.002 &  0.039$\pm$ 0.002 &  0.190$\pm$ 0.003 &  0.099$\pm$ 0.001 & ... &  0.059$\pm$ 0.002\\
  &  1.355 &  0.791$\pm$ 0.002 & ... & ... &  0.513$\pm$ 0.002 &  0.644$\pm$ 0.002 &  0.119$\pm$ 0.002\\
Q0453-0423 &  0.726 &  1.366$\pm$ 0.002 & $<$ 0.51 & $<$ 1.14 &  0.710$\pm$ 0.001 &  1.052$\pm$ 0.001 &  0.455$\pm$ 0.002\\
  &  0.908 &  0.853$\pm$ 0.001 &  0.119$\pm$ 0.002 &  0.373$\pm$ 0.002 &  0.186$\pm$ 0.003 &  0.374$\pm$ 0.002 &  0.119$\pm$ 0.001\\
  &  1.150 &  4.461$\pm$ 0.002 &  1.778$\pm$ 0.002 &  3.492$\pm$ 0.001 &  2.693$\pm$ 0.002 &  3.658$\pm$ 0.001 &  1.524$\pm$ 0.002\\
  &  1.630 &  0.315$\pm$ 0.001 & 0.0041$\pm$0.0008 &  0.015$\pm$ 0.001 & 0.0057$\pm$0.0008 & 0.0052$\pm$0.0005 & ...\\
  &  2.305 &  0.464$\pm$ 0.001 &  0.028 -  0.044 &  0.154$\pm$ 0.001 & ... & ... & $<$ 0.45\\
Q0549-0213 &  0.440 &  0.346$\pm$ 0.007 & ... & ... &   0.24$\pm$  0.02 & $<$  0.22 &  0.034$\pm$ 0.008\\
Q0551-3637 &  1.225 &  0.328$\pm$ 0.004 &  0.006$\pm$ 0.002 &  0.079$\pm$ 0.002 & ... & ... &  0.039$\pm$ 0.003\\
  &  1.961 &   5.15$\pm$  0.03 &  1.732$\pm$ 0.009 &  3.375$\pm$ 0.008 &  2.821 -  3.659 & $<$  3.6 &  1.090$\pm$  0.01\\
Q0926-0201 &  1.106 &  0.375$\pm$ 0.004 & $<$0.0026 &  0.045$\pm$ 0.002 &  0.007$\pm$ 0.001 &  0.038$\pm$ 0.002 &  0.019$\pm$ 0.002\\
Q0940-1050 &  1.789 &  1.129$\pm$ 0.002 &  0.040$\pm$ 0.001 &  0.284$\pm$ 0.001 &  0.095 -  0.111 &  0.280 -  0.300 & ...\\
Q0952+0179 &  0.238 &  1.107$\pm$ 0.006 & ... & ... &   0.65$\pm$  0.01 &   0.84$\pm$  0.01 &  0.361$\pm$ 0.008\\
Q1122-1648 &  0.682 &  1.833$\pm$ 0.002 &  0.399 -  0.441 & 1.2786$\pm$0.0008 &  0.832$\pm$ 0.001 & 1.3125$\pm$0.0008 &  0.140$\pm$ 0.001\\
Q1127-0145 &  0.313 &  1.771$\pm$ 0.005 &   0.89$\pm$  0.02 &   1.18$\pm$  0.01 &  1.204$\pm$ 0.005 &  1.331$\pm$ 0.005 &  1.009$\pm$ 0.005\\
Q1151+0068 &  1.774 &  0.716$\pm$ 0.005 &  0.411$\pm$ 0.005 &  0.546$\pm$ 0.004 &  0.518$\pm$ 0.004 &  0.566$\pm$ 0.004 & ...\\
  &  1.819 &   1.07$\pm$  0.02 &  0.049$\pm$ 0.005 &  0.255$\pm$ 0.005 &  0.105 -  0.235 &  0.192 -  0.268 &  0.138$\pm$ 0.009\\
Q1157+0014 &  1.944 &  1.539$\pm$ 0.006 &  0.882$\pm$ 0.005 &  1.157$\pm$ 0.005 & $<$ 1.1 & $<$ 1.3 &  0.716$\pm$ 0.007\\
Q1202-0725 &  1.755 &  1.899$\pm$ 0.007 & ... & ... &  0.924$\pm$ 0.006 &  1.279$\pm$ 0.006 &  0.180\\ 
Q1229-0021 &  0.395 &  2.064$\pm$ 0.006 & ... & ... &   1.16$\pm$  0.01 &   1.43$\pm$  0.01 &  0.641$\pm$ 0.007\\
Q1246-0217 &  0.727 &  0.453$\pm$ 0.003 &  0.015$\pm$ 0.003 &  0.142$\pm$ 0.005 &  0.060$\pm$ 0.006 & $<$ 0.34 &  0.061$\pm$ 0.003\\
Q1331+0170 &  1.777 &  1.259$\pm$ 0.003 & ... & ... &  0.673$\pm$ 0.002 &  0.823$\pm$ 0.001 &  0.334$\pm$ 0.003\\
  &  1.786 &  0.878$\pm$ 0.003 & ... & ... &  0.311 -  0.369 &  0.505$\pm$ 0.002 &  0.137$\pm$ 0.003\\
Q1337+0113 &  1.637 &  0.617$\pm$ 0.005 &  0.054$\pm$ 0.003 & $<$ 0.30 &  0.156$\pm$ 0.004 &  0.326$\pm$ 0.003 & ...\\
Q1341-1020 &  0.873 &  0.528$\pm$ 0.006 &  0.029$\pm$ 0.005 &  0.208$\pm$ 0.006 &  0.092$\pm$ 0.005 &  0.210$\pm$ 0.004 &  0.078$\pm$ 0.006\\
  &  1.277 &  1.454$\pm$ 0.008 &  0.272$\pm$ 0.007 &  0.802$\pm$ 0.005 &  0.496$\pm$ 0.007 &  0.812$\pm$ 0.005 &  0.303$\pm$ 0.007\\
Q1418-0064 &  1.458 &   2.19$\pm$  0.01 & ... & ... & ... & ... &  0.200$\pm$  0.01\\
  &  1.982 &  0.453$\pm$ 0.008 & $<$0.0058 & ... & $<$0.010 &  0.011$\pm$ 0.002 &  0.013$\pm$ 0.005\\
  &  2.257 &   0.62$\pm$  0.01 & $<$0.0060 & $<$0.0086 & $<$0.011 & $<$ 0.046 & $<$ 0.040\\
Q1444+0014 &  0.660 &  0.341$\pm$ 0.005 & $<$ 0.023 &  0.096$\pm$ 0.003 &  0.022$\pm$ 0.002 &  0.086$\pm$ 0.004 &  0.044$\pm$ 0.002\\
  &  1.159 &  0.722$\pm$ 0.004 &  0.010$\pm$ 0.002 &  0.139$\pm$ 0.003 &  0.040$\pm$ 0.003 & ... &  0.016$\pm$ 0.002\\
Q1621-0042 &  1.133 &  3.168$\pm$ 0.004 & $<$ 1.1 & $<$ 2.3 & $<$ 1.2 & $<$ 3.0 &  0.440$\pm$ 0.005\\
Q1629+0120 &  0.900 &  1.022$\pm$ 0.007 & ... & ... &  0.383$\pm$ 0.007 &   0.70$\pm$  0.01 &  0.205$\pm$ 0.007\\
Q2000-0330 &  2.033 &  0.811$\pm$ 0.002 & $<$ 0.26 &  0.493$\pm$ 0.001 & $<$ 0.35 &  0.490$\pm$ 0.002 & ...\\
  &  2.292 &  0.327$\pm$ 0.003 & 0.0121$\pm$0.0006 & $<$ 0.32 &  0.071$\pm$ 0.003 & ... & $<$ 0.045\\
Q2044-0168 &  1.329 &  0.573$\pm$ 0.005 & ... & ... & ... & ... &  0.032$\pm$ 0.004\\
Q2126-0158 &  2.022 &  0.750$\pm$ 0.003 & $<$ 0.071 & $<$ 0.35 &  0.106$\pm$ 0.001 &  0.282$\pm$ 0.002 & ...\\
Q2206-0199 &  0.752 &  0.885$\pm$ 0.004 & $<$ 0.032 & $<$ 0.20 & ... & ... &  0.232$\pm$ 0.005\\
Q2217-2818 &  0.942 &  0.571$\pm$ 0.001 & 0.0059$\pm$0.0005 &  0.098$\pm$ 0.001 & 0.0307$\pm$0.0006 & 0.0867$\pm$0.0007 & 0.0288$\pm$0.0008\\
  &  1.628 &  0.656$\pm$ 0.001 & 0.0040$\pm$0.0003 & 0.0573$\pm$0.0005 & 0.0239$\pm$0.0007 & 0.0406$\pm$0.0007 & 0.0348$\pm$0.0007\\
  &  1.692 &  1.694$\pm$ 0.001 & 0.0452$\pm$0.0007 & 0.3381$\pm$0.0007 & 0.1159$\pm$0.0009 &  0.346$\pm$ 0.001 & $<$ 0.21\\
Q2225-2258 &  1.412 &  0.418$\pm$ 0.003 & $<$0.0021 &  0.076$\pm$ 0.002 &  0.013$\pm$ 0.001 &  0.052$\pm$ 0.002 & ...\\
  &  1.639 &  0.347$\pm$ 0.002 &  0.004$\pm$ 0.001 &  0.044$\pm$ 0.001 & ... &  0.039$\pm$ 0.002 &  0.021$\pm$ 0.002\\
Q2314-0409 &  1.045 &  0.349$\pm$ 0.005 & $<$0.0064 &  0.033$\pm$ 0.005 & $<$0.0064 &  0.032$\pm$ 0.005 & $<$0.0073\\  
3c336 &  0.318 &   0.48$\pm$  0.01 & ... & ... & ... & ... & $<$0.014\\
  &  0.472 &  0.809$\pm$ 0.009 & ... & ... &   0.11$\pm$  0.01 &   0.26$\pm$  0.01 &  0.035$\pm$ 0.006\\
  &  0.656 &  1.428$\pm$ 0.009 &   0.48$\pm$  0.01 &   1.05$\pm$  0.01 &   0.80$\pm$  0.01 &   1.11$\pm$  0.01 &  0.213$\pm$ 0.009\\
  &  0.797 &   0.46$\pm$  0.01 & $<$0.0089 &  0.029 -  0.091 & $<$0.0083 &  0.028$\pm$ 0.003 & $<$0.0078\\
  &  0.891 &  1.520$\pm$ 0.006 &   0.40$\pm$  0.01 & ... &  0.744$\pm$ 0.007 &  1.114$\pm$ 0.007 &  0.280$\pm$ 0.006\\

\end{longtable}
\begin{minipage}{185mm}
This table includes the measurements of the 87 strong {\MgII} systems detected in our sample of 81 quasar spectra. Col (1): name of the quasar. Col (2): redshift of the absorber. Col (3)-Col(8): measured rest-frame equivalent widths of {\MgII} \lb2796, {\FeII} \lb2374, {\FeII} \lb2383, {\FeII} \lb2587, {\FeII} \lb2600, and {\MgI} \lb2853, respectively. If the transition was covered, but not detected, 3\s upper limits are included after a ``$<$'' sign, while``...'' indicates cases where transitions were not covered in the VLT/UVES spectrum. Ranges of values correspond to lower and upper limits measured in the cases where there was blending. 
\end{minipage}
\end{center}

\clearpage
\newpage

\begin{table*}
\begin{minipage}{185mm}
\begin{center}
\caption{Statistical Results on Very Strong and Moderately Strong Absorption Systems}
\begin{tabular}{l | rrrr}
\hline \hline
 & $\frac{\FeII \lambda2374}{\MgII \lambda2796}$ & $\frac{\FeII \lambda2383}{\MgII \lambda2796}$ & $\frac{\FeII \lambda2587}{\MgII \lambda2796}$ & $\frac{\FeII \lambda2600}{\MgII \lambda2796}$ \\
\hline \hline
$P$(K--S) (moderately vs very strong - all $z_{abs}$) & 1.7e-3 & 0.4e-3 & 7e-6 & 2e-6 \\
$P$(A-D) (same) &   1.6e-3 & 0.05e-3 & 0.02e-3 & 3e-6 \\
\hline
$P$(K--S) (moderately vs very strong -- $z_{abs} < $~1.2) & 1.3e-3 & 0.2e-3  & 3e-6 & 0.6e-06 \\
$P$(A-D) (same) & 0.7e-3 &  0.15e-3 & 0.07e-3 &  4e-06 \\
\hline
$P$(K--S) (moderately vs very strong -- $z_{abs} > $~1.2) & 0.2 & 0.014 &  0.14 & 0.07\\
$P$(A-D)  (same) & 0.14 & 0.009 & 0.08 & 0.09 \\
\hline
$P$ (K--S) (very strong  $z <$~1.2 vs very strong $z_{abs} > $~1.2) & 0.04 & 0.015 & 0.03 & 0.005 \\
$P$ (A-D) (same) & 0.08 & 0.05 & 0.02 & 1.2e-3 \\
\hline
$P$ (K--S) (moderately strong  $z_{abs} <$~1.2 vs moderately strong $z_{abs} > $~1.2) & 0.9 & 0.3 & 0.7 & 0.7 \\
$P$ (A-D) (same) & 0.6 &  0.4 & 0.6 & 0.3 \\
\hline
Bootstrap (moderately strong vs. very strong $z_{abs} <$~1.2) & 0.04\% & 0.8\% & 0\% & 0\% \\
\hline
\end{tabular}
\label{stats}
\end{center}
This table provides the probability ($P$) that several pairs of subsamples of our data points in Figure \ref{ratioz} were derived from the same population. The subsamples were created by selecting different cutoffs of $W_r(\MgII)$ (moderately strong {\MgII} corresponds to 0.3~$< W_r <$~1.0~{\AA} and very strong {\MgII} absorption to $W_r>$~1.0~{\AA}) and redshift (we divide the samples into low and high redshift using the median of the $z_{abs}$ values for each {\FeII} transition and taking the average of the four medians, which corresponds to a value of 1.20). We use the statistical tests Kolmogorov-Smirnov (K--S) and Anderson-Darling (A-D). We include the values for the four {\FeII} transitions analyzed (in each different column). Rows indicate the two subsamples in comparison, except for the last row where we include the results of a bootstrap experiment: we select randomly from the subsample of moderately strong absorbers at low redshift as many points as there are very strong absorbers at low redshift, and calculate the number of times that they cluster at high values of $W_r$(\FeII)/$W_r(\MgII)$ as the subsample of very strong absorbers at low redshift does. Whether or not they cluster was defined as whether the $W_r$(\FeII)/$W_r(\MgII)$ values of the randomly selected data points were larger than the median of the $W_r$(\FeII)/$W_r(\MgII)$ as many times as the values of the very strong absorbers. The value of the median was obtained from all the values of $W_r$(\FeII)/$W_r(\MgII)$ for the very strong and moderately strong subsamples combined; this value differs for each {\FeII} transition. 
\end{minipage}
\end{table*}


\label{lastpage}

\end{document}